\documentclass{emulateapj}
\shorttitle{The IMF of the First stars inferred from EMP stars}
\shortauthors{Ishigaki et al.}

\usepackage{color}
\usepackage{enumerate}
\usepackage{amsmath}
\usepackage{textcomp}
\usepackage{longtable}
\usepackage{ulem}
\usepackage[final]{microtype}
\usepackage{scalefnt}

\usepackage{natbib}
\begin{document}

\title{The Initial mass function of the first stars inferred from extremely metal-poor stars}

\author{Miho N. Ishigaki$^{1}$, Nozomu Tominaga$^{1,2}$, Chiaki Kobayashi$^{1,3}$, and Ken'ichi Nomoto$^{1}$}
\affil{$^{1}$Kavli Institute for the Physics and Mathematics of the Universe (WPI), The University of Tokyo, Kashiwa, Chiba 277-8583, Japan; miho.ishigaki@ipmu.jp}
\affil{$^{2}$Department of Physics, Faculty of Science and Engineering, Konan University, 8-9-1 Okamoto, Kobe, Hyogo 658-8501, Japan}
\affil{$^{3}$School of Physics, Astronomy and Mathematics, Centre for Astrophysics Research, University of Hertfordshire, College Lane, Hatfield AL10 9AB, UK}


\slugcomment{Accepted for publication in the Astrophysical Journal on 23 Feb 2018}

\begin{abstract}
  We compare elemental abundance patterns of $\sim 200$ extremely
  metal-poor (EMP; [Fe/H]$<-3$) stars 
 with supernova
  yields of metal-free stars in order to obtain insights into the
  characteristic masses of the first (Population III or Pop III)
  stars in the Universe.
  Supernova yields are prepared with 
  nucleosynthesis calculations of metal-free stars with 
 various initial masses ($M=$13, 15, 25,
 40 and 100 $M_{\odot}$) and explosion energies ($E_{51}=E/10^{51}$[erg]$=0.5-60$) to include low-energy, normal-energy, and high-energy explosions.
 We adopt the mixing-fallback
  model to take into account possible asymmetry in the supernova
  explosions and the yields that best-fit the observed abundance patterns of
  the EMP stars are searched by varying the model parameters.
  We find that the abundance patterns of the EMP stars
  are predominantly 
  best-fitted with the supernova yields with initial masses
  $M<40 M_{\odot}$, and that more than than half of the stars are
  best fitted with the $M=25 M_\odot$ hypernova ($E_{51}=10$) models.
  The results also indicate
  that the majority of the primordial supernovae
  have ejected $10^{-2}-10^{-1} M_\odot$ of $^{56}$Ni leaving behind
  a compact remnant, either a neutron star or a black hole, with mass
  in a range of $\sim 1.5-5 M_{\odot}$. 
      The results suggest that the masses of the first stars
      responsible for the first
  metal-enrichment are predominantly $< 40 M_{\odot}$. This implies that
  the higher mass first stars were either less abundant or directly
  collapsing into a blackhole without ejecting heavy elements or that a supernova explosion of a higher-mass first star inhibits the formation of the next generation of low-mass stars at [Fe/H]$<-3$. 
\end{abstract}


\section{Introduction} \label{sec:intro}
Nature of
the first (Population III or Pop III) stars 
is crucial in constraining the environment of stars
and galaxy formation in the early Universe. 
The most important characteristic of the Pop III
stars is their typical masses and the 
initial mass function (IMF). 
The masses of the Pop III stars determine
the amount of 
ionizing photons emitted during the stellar evolution and thus they are
important in better quantifying contribution of the Pop III stars
to the cosmic reionization \citep[e.g.,][]{2004ApJ...612..602T}.  
Furthermore, the Pop III stars are responsible for the first metal
enrichment in the Universe, which is one of the important
condition for the formation of the first
low-mass stars \citep{2003Natur.425..812B,2005ApJ...626..627O}.
Thus, amount and composition of elements
synthesized and ejected by the Pop III stars have significant impacts on
subsequent formation of stars and galaxies \citep[e.g.,][]{2011ARA&A..49..373B,2013RvMP...85..809K}.

Previous theoretical studies on the formation of the Pop III stars
based on cosmological simulations suggest that the Pop III stars were
predominantly very massive with the characteristic mass exceeds
$\sim 100$ $M_\odot$ \citep[e.g.,][and references therein]{2004ARA&A..42...79B}. Recent
high-resolution simulations taking into account more detailed physical
processes, however, 
predict that less massive stars with a few tens of $M_{\odot}$ can form
and that the mass ranges extend from subsolar to thousands of $M_{\odot}$ 
\citep{2011Sci...334.1250H,2016ApJ...824..119H,2011ApJ...727..110C,2011ApJ...737...75G,2013MNRAS.433.1094S,2016MNRAS.462.1307S,2013ApJ...773..185S,2014ApJ...792...32S,2014ApJ...781...60H}.
The various mass ranges predicted by the simulations
could partly depend on details of the
simulation techniques and/or numerical resolutions,   
and thus a clear consensus on the
characteristic mass has not been established yet.

Currently, it is not practically feasible to directly observe the Pop III
stars, which are believed to be formed at redshifts of $20-30$.
 Thus, the important
observational probes of the masses of the Pop III stars 
are elemental abundance patterns in long-lived stars so called extremely
metal-poor (EMP; [Fe/H]$<-3$) stars. 
  The Pop III stars are initially formed out of primordial
  (i.e., H and He) gas and produce 
  heavy elements according to their masses and supernova
  explosion energies. The elements 
  are ejected to the interstellar medium via supernovae, whose elemental
  yields depend  on mixing and fallback within
  the progenitor Pop III SNe
\citep{2002ApJ...565..385U,2008ApJ...679..639Z}. The ejecta from the supernova sweep up hydrogen in the
interstellar medium with a certain degree of anisotropy,
which determines the abundances (e.g. [Fe/H]) in
the next-generation stars \citep{1995ApJ...451L..49A,2007ApJ...670....1G,
  2012ApJ...761...56R,2016MNRAS.456.1410S}.
These diluted ejecta may cool and fragment, from which low-mass
stars can form, if the energy injection by the Pop III SNe
has not destructed
the host dark matter halo \citep{2005ApJ...630..675K,2008ApJ...682...49W,2008ApJ...674..644C,2015MNRAS.452.2822S,
  2016MNRAS.463.3354R,2017ApJ...844..111C}.

Given these potentially
complex nature of the heavy-element enrichments by the Pop III SNe, which
could be computationally expensive for numerical simulations to
follow the entire pathways, the atmospheric elemental
abundances in EMP stars have provided unique observational probe
to test theoretical predictions on  
 the physical properties of the Pop III
stars and their supernova explosions.
Systematic searches with
photometry and/or low-to-medium-resolution spectroscopy
have identified a number of EMP star candidates
\citep[e.g.,][and reference therein]{2005ARA&A..43..531B,2015ARA&A..53..631F}.
The follow-up high-resolution spectroscopy has determined
detailed abundance patterns of the EMP stars, which
allow us to study the typical
properties of the first supernovae imprinted on their elemental
abundances \citep{2003Natur.422..871U,2008PASJ...60.1159S,2013ApJ...762...26Y,2013ApJ...778...56C,2013AJ....145...13A,2014AJ....147..136R}.

The following three results have been obtained so far. 
First of all, 
no metal-free stars have been found in the Milky Way so far,
which suggests that the formation of lower-mass ($<0.8 M_\odot$)
Pop III stars, which can survive until
today was suppressed \citep{2015MNRAS.447.3892H,2016ApJ...826....9I,2018MNRAS.473.5308M}. Secondly,
no clear nucleosynthetic signatures of very massive stars
$M\sim 140 -300M_\odot$, such as a very high Si/O ratio resulting
  from a pair-instability supernova \citep{2013ARA&A..51..457N},
have been found \citep[see][for a
  candidate star]{2014Sci...345..912A}. It should be noted, however,
the current surveys may be biased against finding a star
with nucleosynthetic signature of pair-instability Pop III
supernovae because (i) a single pair-instability supernova would
enrich all the gas within its reach to metallicities well
above the EMP surveys, typically targeted at [Fe/H]$<-3$
\citep{2008ApJ...679....6K} and
(ii) pair-instability supernovae are so energetic
\citep[e.g.,][]{2002ApJ...565..385U,2002ApJ...567..532H}
and occur in such low densities \citep[e.g.][]{2004ApJ...610...14W,2004ApJ...613..631K}
that their ejecta escape from the halo and would never been incorporated
into the next-generation of stars.

We also note that extremely massive stars ($300 M_\odot < M < 10^5 M_\odot$) enter the
pair-instability region but continue to undergo gravitational
collapse.  Yields from jet-induced explosions of such stars were
calculated by \citet{2006ApJ...645.1352O} and found to be consistent with the
abundance patterns of intracluster medium but not in good agreement
with EMP stars (e.g., [O/Fe]).

Finally,
the fraction of carbon-enhanced metal-poor (CEMP) stars
increases with decreasing Fe abundance ([Fe/H])
\citep{2013ApJ...762...27Y, 2014ApJ...797...21P}.
The dominance of the CEMP stars among the lowest-[Fe/H]
stars suggests that the characteristic abundance patterns
observed in these stars reflect nucleosynthesis
products of Pop III star's supernova explosions 
\citep{2002ApJ...565..385U,2003Natur.422..871U,2003ApJ...594L.123L,2006A&A...447..623M,2007ApJ...660..516T,2010ApJ...724..341H,2014Natur.506..463K,2014ApJ...792L..32I,2017MNRAS.467.4731C}
and their formation sites.

Interestingly, abundance patterns of EMP stars reported by
previous observational studies \citep[e.g.][]{2004A&A...416.1117C,2008ApJ...681.1524L} are well
explained by nucleosynthetic yields of individual or IMF-averaged
core-collapse supernovae/hypernovae of Pop III stars in a range  
$\sim 10-100 M_\odot$ with
various explosion energies
\citep{2005Sci...309..451I,2007ApJ...660..516T,2008ApJ...681.1524L,2010ApJ...709...11J,2010ApJ...724..341H,2014ApJ...785...98T,2015ApJ...809..136P}. 
The 2D numerical simulations of Pop III supernovae with progenitor masses
in a range $15-40M_\odot$ by \citet{2007ApJ...657L..77T,2009ApJ...690..526T,2010ApJ...709...11J} also reasonably
well reproduce the observed elemental abundances.
  Given the growing observational data for the EMP stars,
  the constraints on the Pop III star's
  masses should be statistically studied based on larger samples.

It is not straightforward, however, to accurately
predict nucleosynthesis products finally ejected by
the core-collapse supernovae of Pop III stars, namely, supernova yields.
The main reason for the uncertainty in the supernova yields is that
the explosions could be highly non-spherical as evidenced by both
observations \citep[e.g.,][]{2008Sci...319.1220M} and theoretical calculations
\citep[e.g.,][]{2012ARNPS..62..407J,2012PTEP.2012aA301K,2013RvMP...85..245B,2015A&A...577A..48W}.
While the ejecta is determined by the progenitor
star structure and the explosion energy in the case of a spherical explosion,
multi-dimensional calculations of mixing and fallback of the ejecta
are needed to take into account the non-sphericity \citep{2009ApJ...693.1780J,2010ApJ...723..353J}.

In order to approximately take into account the effects of
the aspherical explosion in
calculating supernova yields, the mixing-fallback model has been proposed
by \citet{2002ApJ...565..385U}. The  mixing-fallback model mimics an aspherical
explosion with three parameters and is 
adopted on one-dimensional nucleosynthesis calculations of
Pop III supernovae \citep[see Appendix of][]{2007ApJ...660..516T}.
The resulting yields obtained from the model well reproduce the characteristic
nucleosynthesis yields
of aspherical explosions from two-dimensional
simulations \citep[e.g.][]{2009ApJ...690..526T,2010ApJ...709...11J,2017MNRAS.467.4731C} and 
successfully 
explain the key elemental abundances observed in EMP stars
\citep{2002ApJ...565..385U,2003Natur.422..871U,2005ApJ...619..427U,2005Sci...309..451I,2007ApJ...660..516T,2014ApJ...785...98T,2014ApJ...785L...5K,2014ApJ...792L..32I}.

Given that the actual mechanisms of supernova explosions have not yet been
well established \citep[e.g.,][]{2012ARNPS..62..407J}, the mixing-fallback model allows us to
explore the parameter spaces that cover a wide range of
mixing and fallback with much less computational cost
than the multi-dimensional
simulations. This enables us to obtain  
the typical properties of 
the Pop III stars by fitting the abundances of
large statistical samples of EMP stars (see Section \ref{sec:M-F}).
The model also provides a framework to empirically constrain
the degree of asymmetry, the ejected mass of radioactive $^{56}$Ni, which
powers the supernova lightcurve, and the mass of
the compact remnant, either
a neutron star or a black hole, left behind the Pop III supernovae.

In this paper, we calculate a grid of the supernova yield sets of Pop III stars
using the mixing-fallback model, and determine the best-fit model to reproduce each
elemental abundance pattern in
$\sim 200$ EMP stars compiled from the recent literature. By applying
the abundance fitting method to the large sample of EMP stars, 
 we obtain the distributions of 
 mass, explosion energy and the state of mixing and fallback of the
 Pop III supernova models. 
 In the abundance fitting analysis, we take into account
 the theoretical uncertainties arising from stellar evolution
 and the supernova explosion mechanisms. We also examine the effects of
 observational uncertainties on the best-fit models as well as on
 the inferred IMF of the Pop III stars. Based on the obtained best-fit
 models, we discuss the diagnostic elemental abundances that
 are sensitive to the Pop III masses, which will be useful in
 interpreting data from on-going and future spectroscopic surveys of EMP stars.

This paper is organized as follows.
In Section \ref{sec:method}, we first describe our method to
calculate the Pop III supernova yields.  
Then, we describe the observational data of EMP stars from literature
in Section \ref{sec:obsdata}. 
The results of the abundance fitting, the effects of observational
uncertainties, and the comparison with literature
are presented in Section \ref{sec:results} and their
implications are described in Section \ref{sec:discussion}.
Finally, we present the summary of the present analyses in Section \ref{sec:conclusion}.

\section{Abundance fitting method} \label{sec:method}

\subsection{Supernova yields}

We obtain nucleosynthesis yields of Pop III stars 
making use of progenitor models and explosive nucleosynthesis
previously calculated by \citet{2000fist.conf..150U}, \citet{2005Sci...309..451I}, \citet{2005ApJ...619..427U}, and \citet{2007ApJ...660..516T}. 
The progenitor models were calculated through Fe core collapse
for the initial masses of 13, 15, 25, 40, and 100$M_\odot$.
The Henyey-type stellar evolution code was used 
\citep[][and reference therein]{1988PhR...163...13N,2000fist.conf..150U} with a nuclear reaction
network as in \citet{1996ApJ...460..869H}.
The abundance ratios of C, O, Ne, Mg, and Al in the core
are largely influenced  by the uncertain
$^{12}$C($\alpha$,$\gamma$)$^{16}$O reaction
rate \citep{1984RvMP...56..149F,2002ApJ...577..281C}, for which 
1.4 times the value given in \citet{1988ADNDT..40..283C} was adopted.

The explosive nucleosynthesis was calculated by injecting
the thermal energy in an innermost region of the progenitor \citep{2007ApJ...660..516T}.
   For the explosive
   burning, a reaction network including 280 species up to $^{79}$Br is used
   as in \citet{2005ApJ...619..427U}. 

We adopt nine pairs of
progenitor initial masses (13, 15, 25, 40, and 100$M_{\odot}$) and
explosion energies (normal supernovae with $E_{51}=E/10^{51} {\rm erg}=1$,
low-energy supernovae with $E_{51}=0.5$, 
and hypernovae with $E_{51}\ge 10$), as summarized in Table \ref{tab:snmodels}.
In the following, we denote these models as 13LE for the low-energy supernova
with the progenitor initial mass $M=$13$M_{\odot}$, 
13SN, 15SN, 25SN, 40SN, and 100SN for normal-energy
supernovae with progenitor masses
$M=$13, 15, 25, 40, and 100$M_{\odot}$, respectively, and 25HN, 40HN, 100HN for hypernovae with progenitor masses $M=$25, 40, and 100$M_{\odot}$, respectively.

  Among these massive stars, the $100M_\odot$ star undergoes
  pulsational pair-instability (PPI) and eventually Fe core collapse,
  which is common for $80-140 M_\odot$ \citep[e.g.][]{2002ApJ...567..532H}.
  One example of such an evolutionary track of the central
  density and temperature is seen in Figure 7 of \citet{2009ApJ...706.1184O}
  for the Pop III 135 $M_\odot$ star. Before PPI, the 135 $M_\odot$
  star has much higher central entropy compared with the 40 $M_\odot$
  star at similar nuclear burning stages. PPI, however, delays
  the onset of Fe core collapse. As a result, the central entropy of
  the 135 $M_\odot$ star is decreased by neutrino emissions during
  PPI and becomes as low as that of the 40 $M_\odot$ star at the
  beginning of Fe core collapse (Fig. 7 of \citet{2009ApJ...706.1184O}). The
  hydrodynamical behavior of collapse of such a low entropy
  Fe core after PPI has not been well studied.
  The supernova explosion and nucleosynthesis from such stars
  may not necessarily by only HN-like but cold also be normal-energy
  SN-like. Since there is no other empirical probe of the zero-metallicity
  supernova events and recent cosmological simulations \citep[e.g.][]{2014ApJ...781...60H}
  predict the formation of $\sim 100 M_\odot$ Pop III stars, we
  include not only 100HN but also 100SN to investigate whether
  or not signature of such star is found in EMP stars.
For each model, we apply the mixing-fallback model to calculate
the supernova yields as described in the next subsection.

\subsection{The mixing-fallback model \label{sec:M-F}}

In order to take into account the non-sphericity in
supernova explosions, we apply the mixing-fallback
model as adopted in \citet{2002ApJ...565..385U,2005ApJ...619..427U,2007ApJ...660..516T}. In this model, 
mixing of ejecta and the amount of fallback are described
by the three parameters, the initial mass cut $M_{\rm cut}$,  the outer boundary
of mixing $M_{\rm mix}$, and the ejection fraction $f_{\rm ej}$.
The $M_{\rm cut}$ represents the boundary, above which the
nucleosynthesis products can potentially be ejected. The $M_{\rm mix}$ represents
the outer boundary of the mixing zone, above which all materials
are ejected. The fraction $f_{\rm ej}$ of the material contained in
the mixing zone (i.e., the layers between $M_{\rm cut}$ and $M_{\rm mix}$)
is finally ejected to interstellar medium, while the remaining
material falls back to the central compact remnant.

In our abundance fitting procedure,
we fix $M_{\rm cut}$ at the surface of the Fe core,
where the mass fraction of $^{56}$Fe 
dominates over that of $^{28}$Si in the pre-supernova progenitor, 
while $M_{\rm mix}$ and $f_{\rm ej}$ are free parameters. 
The adopted model properties are summarized in Table \ref{tab:snmodels}.

The outer boundary of mixing ($M_{\rm mix}$) is varied
  as a function of $x$,
  where $M_{\rm mix}=M_{\rm cut}+x(M_{\rm CO}-M_{\rm cut})$, in the range
  of $x=0.0-2.0$ with a stop of $0.1$. The range of $M_{\rm mix}$ is
  thus from $M_{\rm cut}$ up to 
  $M_{\rm cut}+2.0(M_{\rm CO}-M_{\rm cut})$.
The ejection fraction $f_{\rm ej}$ is logarithmically varied
from $\log f_{\rm ej}=-7$ to $0$ with a step of 0.1 dex.
Based on the best-fit parameters, we also calculate the 
mass of the compact remnant left behind following
the fallback based on the following relation from \citep{2007ApJ...660..516T}:

\begin{equation}
  M_{\rm rem}=M_{\rm cut}+(1-f_{\rm ej})(M_{\rm mix}-M_{\rm cut})
  \label{eq:mrem}
\end{equation}

\begin{deluxetable*}{lcccccc}
\tablecaption{Properties of the models \label{tab:snmodels}}
\tablehead{
\colhead{Model ID} & \colhead{$M$\tablenotemark{a}} & \colhead{$E$\tablenotemark{b}} & \colhead{$M_{\rm cut}$\tablenotemark{c}} & \colhead{$M_{\rm CO}$\tablenotemark{d}} & \colhead{Maximum $M_{\rm mix}$\tablenotemark{e}} & \colhead{Range of $M(^{56}$Ni$)$\tablenotemark{f}}\\
\colhead{} & \colhead{($M_\odot$)} & \colhead{($10^{51}$ erg)} & \colhead{($M_\odot$)} & \colhead{($M_\odot$)} & \colhead{($M_\odot$)} & \colhead{($M_\odot$)}
}
\startdata
13LE & 13 & 0.5 & 1.47 & 2.39 & $3.30$ & 1.55$\times 10^{-8}-10^{-1}$ \\
13SN & 13 & 1 & 1.47 & 2.39 & $3.30$ & 1.67$\times 10^{-8}-10^{-1}$\\
15SN & 15 & 1 & 1.41 & 3.02 & $4.64$ & 1.32$\times 10^{-8}-10^{-1}$\\
25SN & 25 & 1 & 1.69 & 6.29 & $10.90$ & 2.61$\times 10^{-8}-10^{-1}$\\
25HN & 25 & 10 & 1.69 & 6.29 & $10.90$ & 6.72$\times 10^{-8}-10^{-1}$\\
40SN & 40 & 1 & 2.42 & 13.89 & $25.36$ & 5.18$\times 10^{-8}-10^{-1}$\\
40HN & 40 & 30 & 2.42 & 13.89 & $25.36$ & 1.96$\times 10^{-7}-10^{0}$\\
100SN & 100 & 1 & 3.63 & 42.00 & $80.37$ & 4.93$\times 10^{-7}-10^{0}$\\
100HN & 100 & 60 & 3.63 & 42.00 & $80.37$& 8.13$\times 10^{-7}-10^{}$ \\
\enddata
\tablenotetext{a}{Progenitor mass.}
\tablenotetext{b}{Explosion energy.}
\tablenotetext{c}{Initial mass cut, which corresponds
  to the lower bound for the considered $M_{\rm mix}$ range.}
\tablenotetext{d}{CO-core mass.}
\tablenotetext{e}{Upper bound for the $M_{\rm mix}$ range.}
\tablenotetext{f}{The range in ejected mass of $^{56}$Ni for the parameter space considered in each model.}
\end{deluxetable*}

\subsection{Abundance fitting procedure}

From the grid of SN yields for the nine models with varying
parameters in Table \ref{tab:snmodels}, 
the best-fit models
are searched by minimizing
$\chi^2_{\nu}=\chi^2/\nu$, where the degree of freedom,
$\nu$, refers to the value 
$\nu=N-M$, where $N$ and $M$ are the number of abundance data
points and the number of parameters (mass, energy,
$M_{\rm mix}$, $f_{\rm ej}$ and the hydrogen mass), respectively.  The $\chi^2$ is
defined below similar to that employed in \citet{2010ApJ...724..341H},

\begin{eqnarray}
  \chi^{2}&=&\sum\limits_{i=1}^{N}\frac{(D_{i}-M_{i})^2}{\sigma_{o,i}^2+\sigma_{t,i}^2}\nonumber\\
  &+&\sum\limits_{i=N+1}^{N+U}\frac{(D_i-M_i)^2}{(\sigma_{o,i}^2+\sigma_{t,i}^2)}\Theta(M_i-D_i) \nonumber \\
 &+&\sum\limits_{i=N+U+L}^{N+U+L}\frac{(D_i-M_i)^2}{(\sigma_{o,i}^2+\sigma_{t,i}^2)}\Theta(D_i-M_i)
\end{eqnarray}

where $D_i$ and $M_i$ are observed and model values of [X/H], respectively, and for an element $i$,
$\sigma_{o,i}$ and $\sigma_{t,i}$ are corresponding to observational and theoretical uncertainties, respectively. 
The Heaviside function $\Theta(x)$ is defined to be $\Theta(x)=1$ for $x>0$ and $0$ otherwise.
The observational upper (lower) limits are only taken into account if these limits are below (above) the model values via the second (third) term in the above expression.
The theoretical lower limit described in \ref{sec:theoryerr},
is implemented in the second term with the same expression as the
observational upper limit. In this analysis, the hydrogen mass
to calculate [X/H] abundance in the model is also varied as a free parameter
to take into account a wide range of [Fe/H] for the second-generation
stars predicted by 
hydrodynamical simulations \citep[e.g.,][]{2012ApJ...761...56R}.

For C and N abundances, we fit the combined value [(C+N)/H] rather than
treating [C/H] and [N/H] separately because of the two reasons;
First, atmospheric abundances of evolved EMP stars might have been
affected by internal mixing, by which material from the H-burning shell
is dredged up. Since C is processed into N in the H-burning shell
via CNO cycle, the surface abundances with the internal mixing would have
been enhanced in N in expense of C. Second, the above process could have
also occurred within
the progenitor Pop III star before its supernova explosion. 
Therefore, the combined C$+$N abundance is not significantly affected, although  
the individual C and N abundances
may have changed from the original value.

\subsection{Theoretical uncertainties \label{sec:theoryerr}}

The theoretical uncertainties stem from physical mechanisms
that are treated approximately (overshooting, etc.) or are not
taken into account (stellar rotation and/or $\nu$-process, etc.) in the
employed model. We discuss these limitation in more detail
in Section \ref{sec:limitation}.

Among the elements mainly produced during the stellar evolution,
Na and Al are known to be subject of several uncertainties.
Since these elements are synthesized in the C-shell burning
via the reaction $^{12}$C($^{12}$C, $p$)$^{23}$Na($\alpha$,$\gamma$)$^{27}$Al,
the Na and Al abundances are
sensitive to the $^{12}$C abundance after core He burning
and the temperature of the C-shell burning. They depend on
the $^{12}$C($\alpha$,$\gamma$)$^{16}$O reaction
rate \citep{2002ApJ...577..281C} and the overshooting \citep{2005Sci...309..451I}.
For these reasons, 
we assume a larger theoretical uncertainty ($\sigma_{t,i}$)
of 0.5 dex for Na and Al. 

Titanium and Scandium are known to be under produced in one-dimensional
calculations of supernova nucleosynthesis compared to those observed in
EMP stars \citep[e.g.,][]{2007ApJ...660..516T,2016ApJ...817...53S}. Several possible sites have
been proposed for their synthesis such as the neutrino process
\citep{2011ApJ...739L..57K} and/or
the jet-induced explosions \citep[][and references therein]{2009ApJ...690..526T}. Since the main production sites have not been clearly identified
because of the uncertainties in the physical mechanisms of supernovae,
we treat the model abundances of Ti and Sc as lower limits.
For the other elements, the theoretical uncertainties are assumed to be zero.

\section{Abundance data from literature} \label{sec:obsdata}

We employ the elemental abundance data available from
recent studies for large samples of EMP stars
\citep{2013ApJ...762...26Y,2013ApJ...778...56C,2014AJ....147..136R,2015ApJ...807..171J} and newly identified UMP stars \citep{2014ApJ...787..162H,2015ApJ...809..136P,2015ApJ...810L..27F,2016AA...585L...5M,2016ApJ...833...21P}.
These studies are selected so that the abundance measurements
of C, N, O, Na, Mg, Al, Si, Ca, Sc, Ti,
Cr, Mn, Fe, Co, Ni, and Zn have been performed
based on high-resolution spectra with a
spectral resolution greater than $R\sim 28000$. 
Some EMP stars are analyzed in more than one references,
for which we take the data from the study with the largest number of
measured elemental abundances.

In the following analysis, we restrict our sample to EMP ([Fe/H]$<-3$) stars,
which were presumably formed out of gas predominantly enriched by a single
supernova. With the [Fe/H] criterion, 
the number of unique stars is 219. To examine whether a star is
likely polluted by an evolved AGB binary companion, we check their
Sr and Ba abundances. Three stars in the sample have [Sr/Fe]$>1$
but none of them show [Ba/Fe]$>1$.
Therefore, we keep these stars in our sample.

In the abundance fitting analysis, we consider abundances relative to
the solar abundance of \citet{2009ARA&A..47..481A} ([X/H]) of C$+$N, O, Na, Mg, Al, Si, Ca, Sc, Ti, Cr, Mn, Fe, Co, Ni, and Zn.
Among these elemental abundances, [Fe/H] abundances are measured
from the largest number of absorption lines and thus we assume 
the smallest error of 0.1 dex in the abundance fitting analysis. 
In most of the literature data, C and N abundances have been
measured from spectral fitting of CH and CN
molecular features and are known to be sensitive to the 3D effects
in the stellar atmosphere \citep[e.g.,][]{2016A&A...593A..48G}.
In this paper, we adopt 0.2 and 0.3 dex for the uncertainties of
[C/H] and [N/H], respectively. 
Significant non-LTE effects 
are expected for O abundances
measured from IR OI triplet lines at 777 nm, Na abundances
from Na I D doublet and Al abundances from Al I 3961 {\AA} line \citep{2009A&A...500.1221F,2011A&A...528A.103L,1997A&A...325.1088B}, which are adopted in most of these studies.
In particular, the non-LTE (NLTE)
correction to the Al abundances ranges from $\sim 0.2$ dex up to $\sim 0.6$ dex
for EMP stars depending
on stellar atmospheric parameters. In the following analysis, except for
the three stars with [Fe/H]$<-5$ in Section \ref{sec:HMP},
we use uncorrected values for the O, N, and Al abundances but assign
a relatively large error of 0.3 dex for these elements. 
In most of the studies Si abundances are determined from
only one or a few lines, and we adopt
a larger error of 0.2 dex. 
The large errors of 0.2 dex are also assigned to the Ti and Cr abundances,
since their abundances from neutral and ionized species generally disagree
by 0.1-0.5 dex in 1D-LTE analyses \citep{2006ApJ...653.1145K,2014AJ....147..136R}.
A relatively large error of 0.2 dex is also assigned to Mn abundances
since they are most frequently measured from the
resonance  lines at $\sim 4030$ {\AA} and the derived Mn abundances 
have been reported to disagree with the
values from non-resonance lines by $\sim 0.3$ dex \citep[e.g.,][]{2013ApJ...778...56C}. 
For the other elements, an observational uncertainty of 0.15
dex is assumed. In the following analysis, we restrict our sample to
those having observed [X/H] constraints, excluding upper limits, greater
than seven to make sure that the 5 model parameters (mass, energy,
$M_{\rm mix}$, $f_{\rm ej}$, and hydrogen mass) are well constrained.
We differ our discussion on the three most Fe-poor stars, HE 0107-5240
\citep{2002Natur.419..904C}, HE1327-2326 \citep{2005Natur.434..871F,2006ApJ...639..897A}, and SMSS 0313-6708 \citep{2014Natur.506..463K}, based on the most recent abundance measurements for these
stars in Section \ref{sec:HMP}.

\section{Results} \label{sec:results}

As a result of the abundance fitting described in the previous sections, 
the $M=15M_{\odot}$/supernova (15SN), $25M_{\odot}$/supernova (25SN), $25M_{\odot}$/hypernova (25HN), $40M_{\odot}$/hypernova (40HN), or $100M_{\odot}$/supernova (100SN) models
best-fit with $\chi^2_{\nu}<3$ the abundance patterns of at least one EMP star.
The best-fit 13$M_\odot$/low-energy (13LE) or 100$M_{\odot}$/
hypernova (100HN) models are also found but they result in
larger $\chi^2_{\nu}$ values.

The best-fits were
not found for the 13SN or 40SN models.
The observed abundances ([Fe/H] and [X/Fe])
and their best-fit models (one of the 15SN, 25SN, 25HN, 40HN, or
100SN models)
are summarized in Figure \ref{fig:feh_xfe} and 
Table \ref{tab:bestfitmodels} in the
Appendix (published entirely in an electronic form).

In our analysis, we have checked whether there is only
one best-fit peak or there are more than one local $\chi^2$ minima
in the parameter space. 
Figures \ref{fig:chi2dist} and \ref{fig:chi2distcemp} show the
example plots, where we plot
the behavior of a $p$-value, which is calculated
as an integral of a $\chi^2$ probability
distribution of a given degree of freedom above the observed $\chi^{2}$
value, in the planes of the model parameters.
The nine different panels correspond to the
nine models (Table \ref{tab:snmodels})
considered in this work. Each panel plots the $p$-values by colors in the
$\log f_{\rm ej}$-$x$ space, where $x$ is the scale factor for $M_{\rm mix}$
(see Section \ref{sec:M-F} definition).

We confirm that the region with the largest $p$-values
is found only around the best-fit parameter
(shown by a cross) and no secondary minimum are found
for the $\log f_{\rm ej}-x$ parameter space of
a given model. The distinction between the different
models are, however, not always clear.
For example, for the case of CS 22941-017
(Figure \ref{fig:chi2dist}), 
 the region with larger $p$-values is
located at $\log f_{\rm ej}>-2$ and
$x<0.5$ not only in the best-fit model (25HN) but also in the 
other models such as the 40HN model. For such a case with
multiple models that have similar $p$-values, we show
our results with and without $p$-value weighted.
In addition, we investigate the effects of
observational and theoretical uncertainties on the resulting
inference on the masses of the models in Section \ref{sec:testerr}.  
For the case of CS 29498-043 (Figure \ref{fig:chi2distcemp}), for which
all of the abundance measurements considered in the present analysis are available,
higher-$p$-values are only seen around the best-fit
parameters in the 25HN model.

To illustrate the characteristic abundance patterns of the models
of the different masses and the explosion energies,
Figure \ref{fig:representative} shows
the best-fit models (15SN, 25SN, 25HN, 40HN, and 100SN from top to bottom)
and the observed abundances for
the stars fitted with relatively small $\chi^2_{\nu}$.
It can be seen that the odd-even effect among
Na, Mg, Al and Si abundances are
stronger in the $M=25M_{\odot}$/supernova model than
in the $M=15M_{\odot}$/supernova model.

In the following subsections we describe the [X/Fe]
ratios of the best-fit models for each of the 15SN,
25SN, 25HN, 40HN, and 100SN models.

\begin{figure*}
  \begin{center}
    \includegraphics[width=18cm]{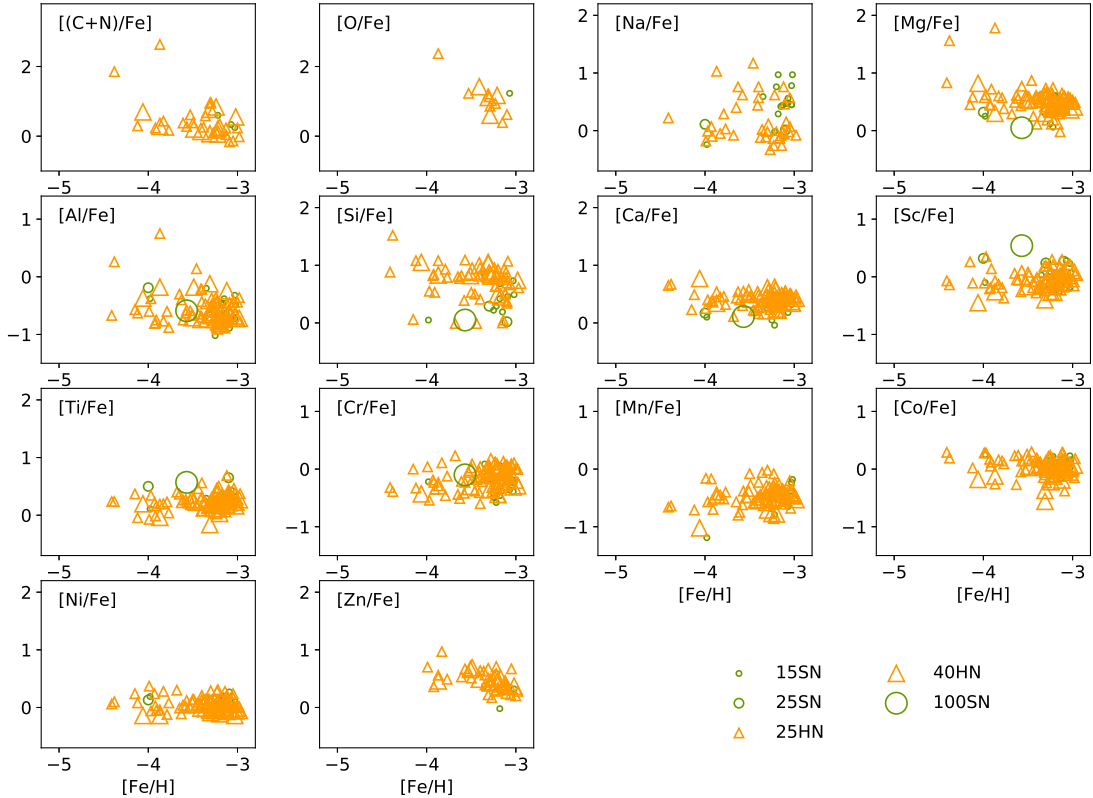}
\end{center}
\caption{Observed abundance ratios ([X/Fe]) of the present sample of EMP stars
  that are fitted by the models with $\chi^2_{\nu}<3$, plotted
  against [Fe/H]. Symbols indicate the best-fit
  models (either 15SN, 25SN, 25HN, 40HN, or 100SN models) for individual
  stars. Meaning of the symbols is shown on the lower-right corner. \label{fig:feh_xfe}}
  \end{figure*}

\begin{figure*}
  \plotone{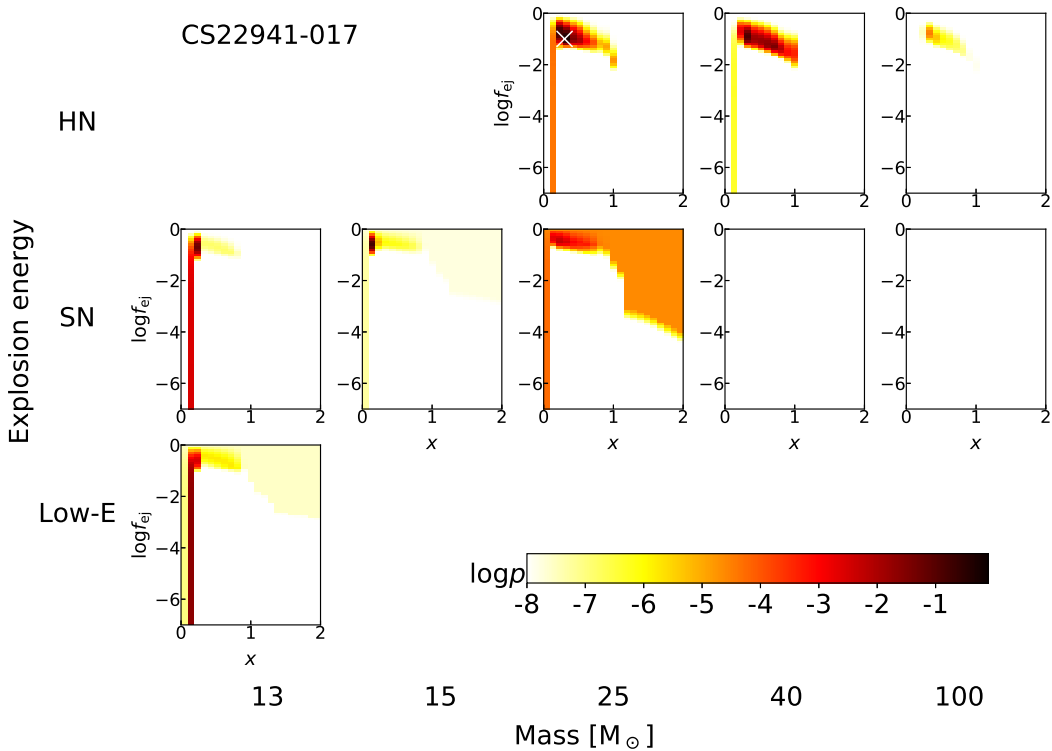}
  \caption{Distribution of the $p$-values calculated as an integral of
    $\chi^2$ probability distribution for a given degree of freedom
    in the parameter spaces for one of the EMP stars, CS22941-017.
    Different panels correspond to the models with various 
    progenitor masses (increasing along the horizontal axis) and explosion
    energies (increasing along the vertical axis; from bottom to top, low-energy or ``LE'', normal-energy or ``SN'', and high-energy or ``HN'' explosions).
    Each panel shows the $p$-values in
    a $\log f_{\rm ej}$-$x$ plane, where $x$ is the scale factor of $M_{\rm mix}$ (see Section \ref{sec:M-F}). The location marked by a white $x$ indicates
    the best-fit parameters.\label{fig:chi2dist}}
\end{figure*}

\begin{figure*}
  \plotone{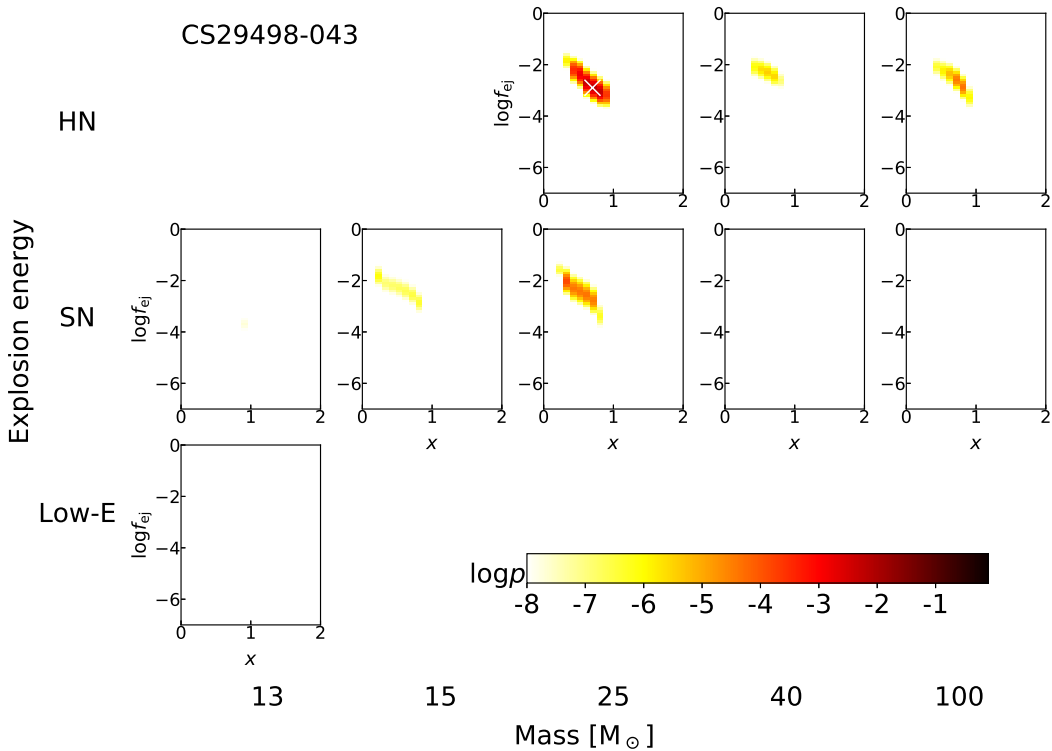}
  \caption{Same as Figure \ref{fig:chi2dist} but for a CEMP star, CS 29498-043.\label{fig:chi2distcemp}}
\end{figure*}

\begin{figure*}
\plotone{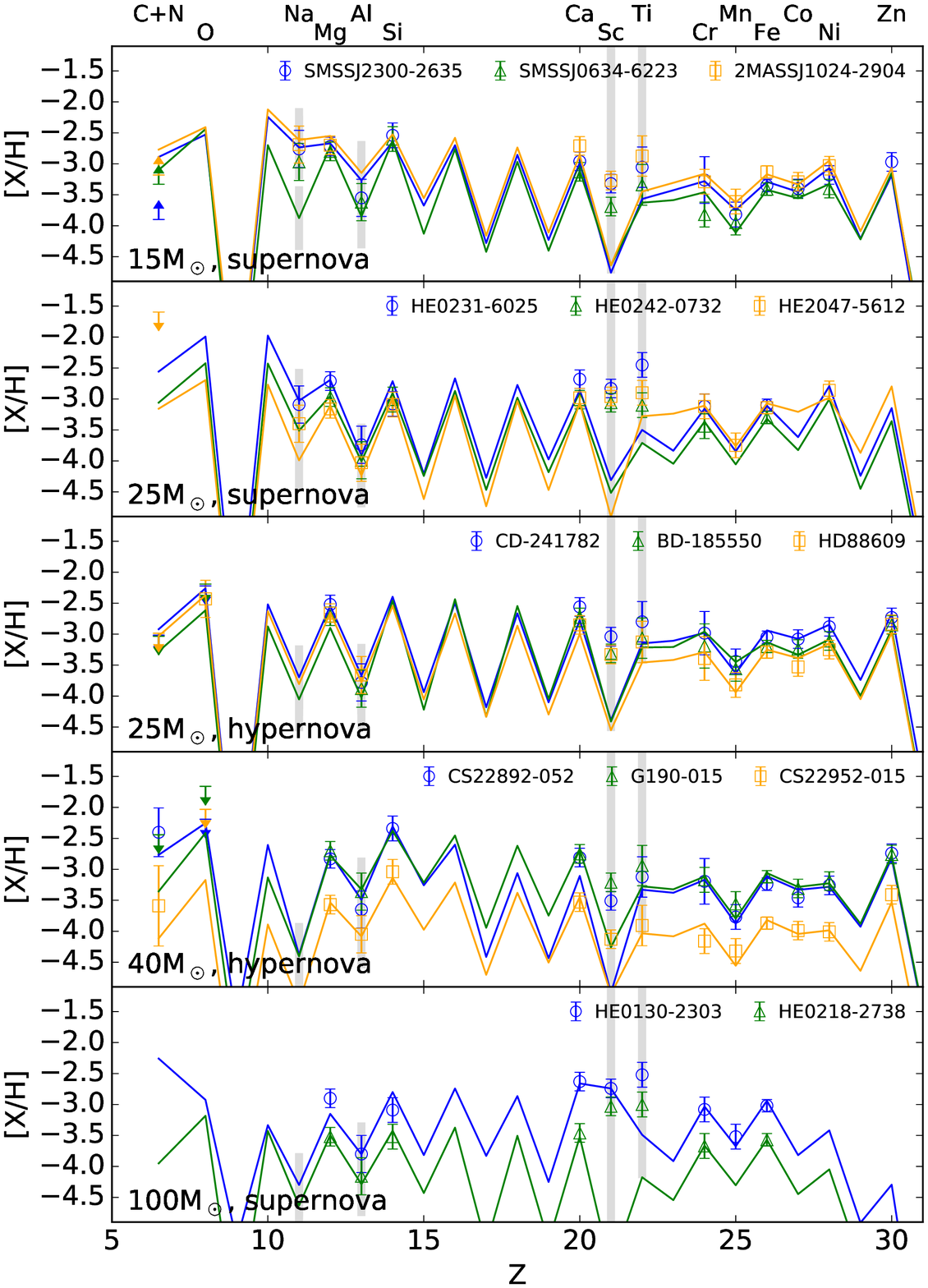}
\caption{ The observed abundances (circle, triangle and square
  with error bars, arrows
  for upper limits) and the best-fit models (solid lines with color
  corresponding to each star) for stars with
  relatively small $\chi^{2}_{\nu}$ among those fitted with each of the
  five models. From top to bottom, the stars best-fitted
  with the 15SN, 25SN,
  25HN, 40HN, and 100SN models, respectively, are shown. The
  model abundances of elements
  marked by gray bars are either assigned a large theoretical uncertainty
  (Na and Al) or treated as a lower limit (Sc and Ti).
    \label{fig:representative}}
\end{figure*}

\subsection{Characteristic abundance ratios\label{sec:representative}}

\begin{figure*}
  \begin{center}
    \begin{tabular}{c}
      \begin{minipage}{0.5\hsize}
        \begin{center}
          \includegraphics[width=9.0cm]{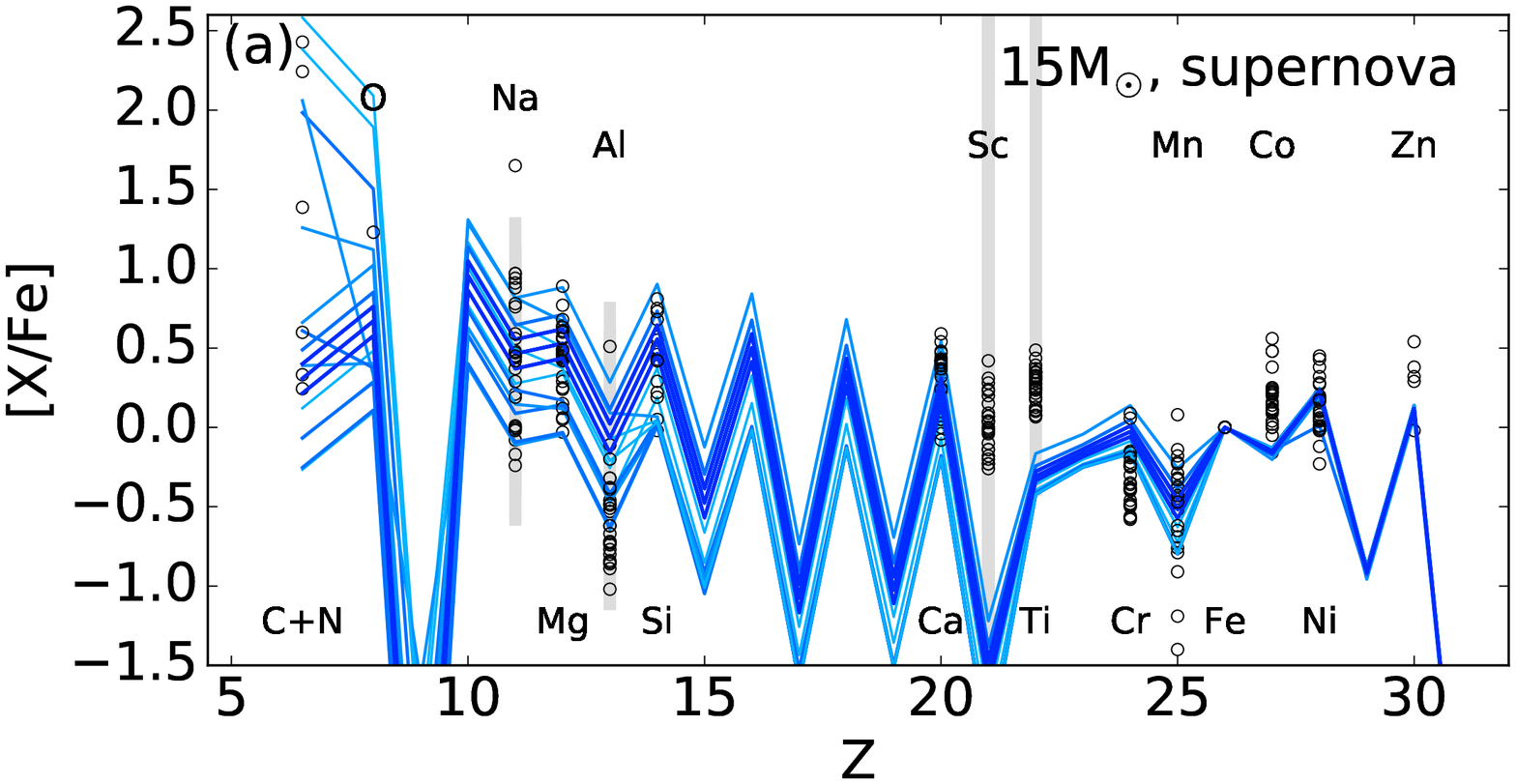}
        \end{center}
      \end{minipage}
      \begin{minipage}{0.5\hsize}
        \begin{center}
          \includegraphics[width=7.7cm]{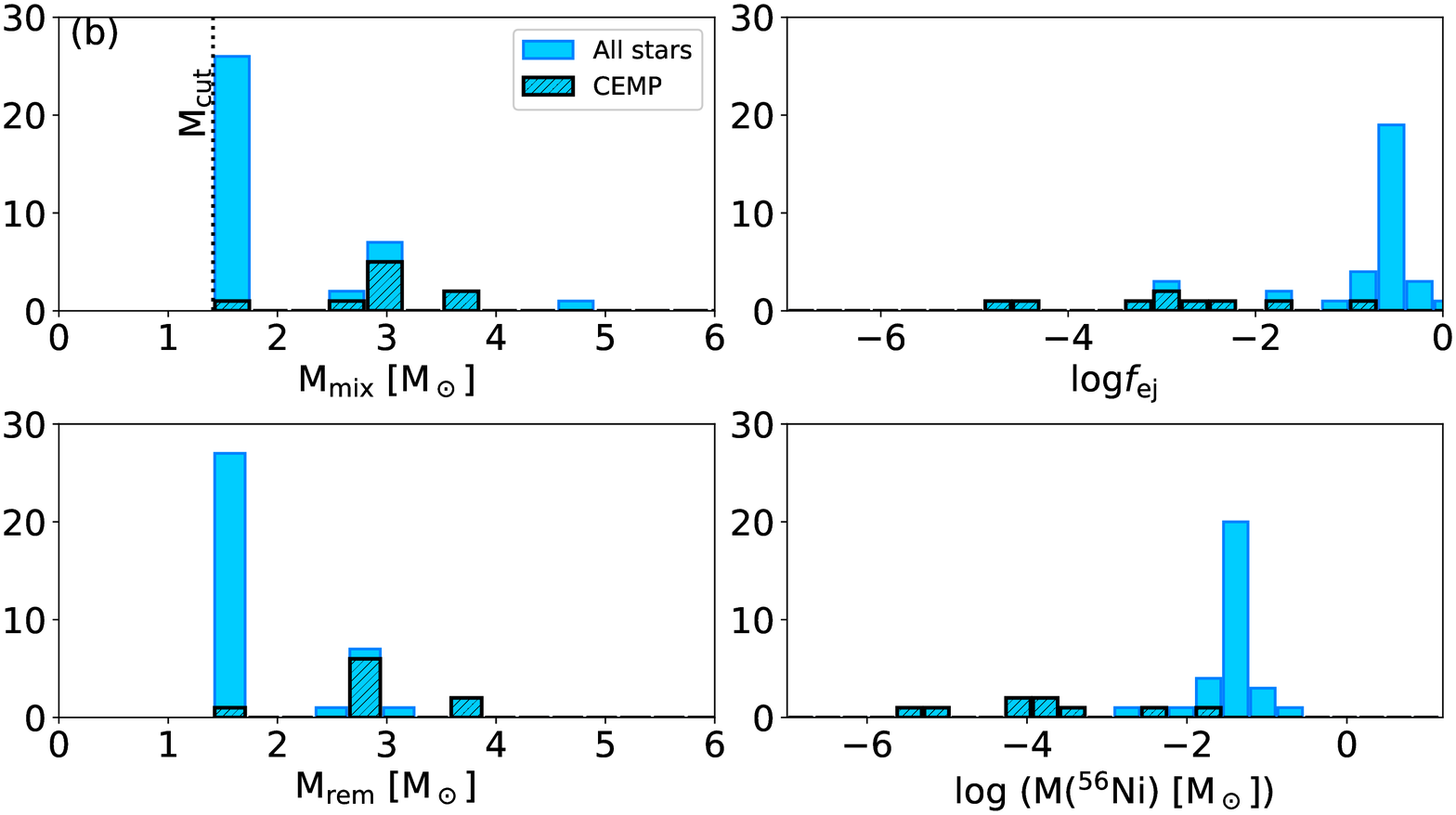}
        \end{center}
      \end{minipage}\\
      \begin{minipage}{0.5\hsize}
        \begin{center}
          \includegraphics[width=9.0cm]{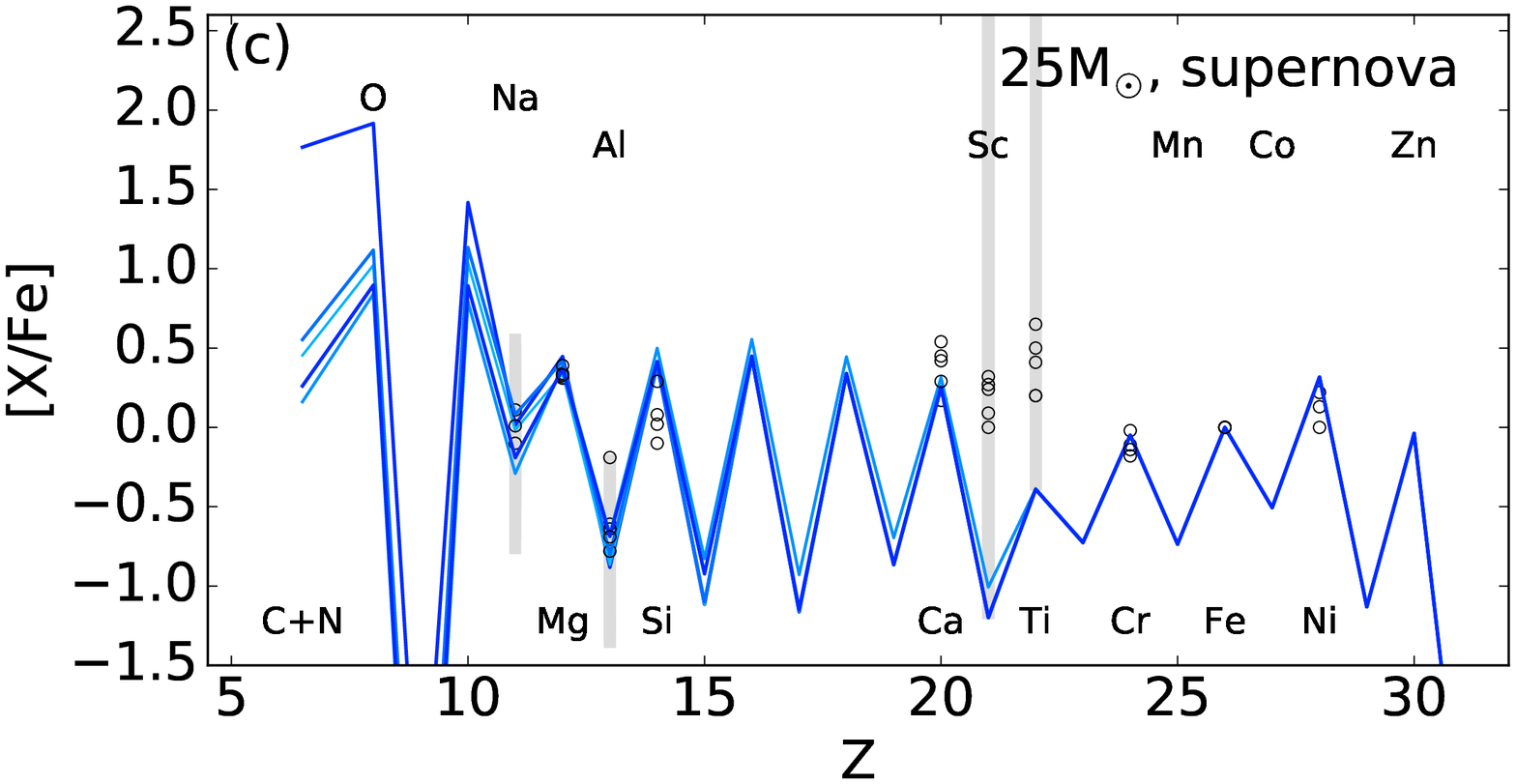}
        \end{center}
      \end{minipage}
      \begin{minipage}{0.5\hsize}
        \begin{center}
          \includegraphics[width=7.7cm]{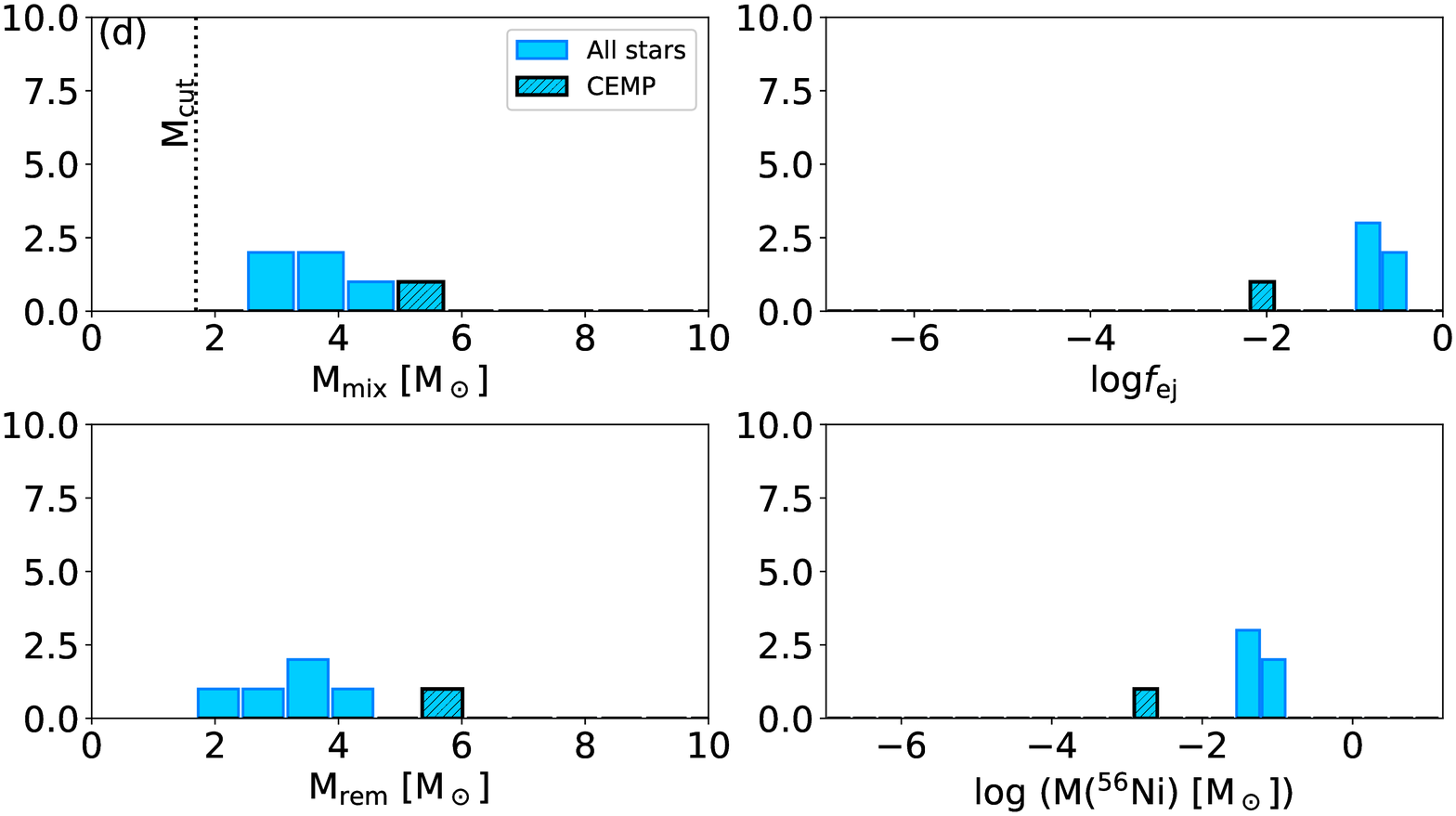}
        \end{center}
      \end{minipage}\\
      \begin{minipage}{0.5\hsize}
        \begin{center}
          \includegraphics[width=9.0cm]{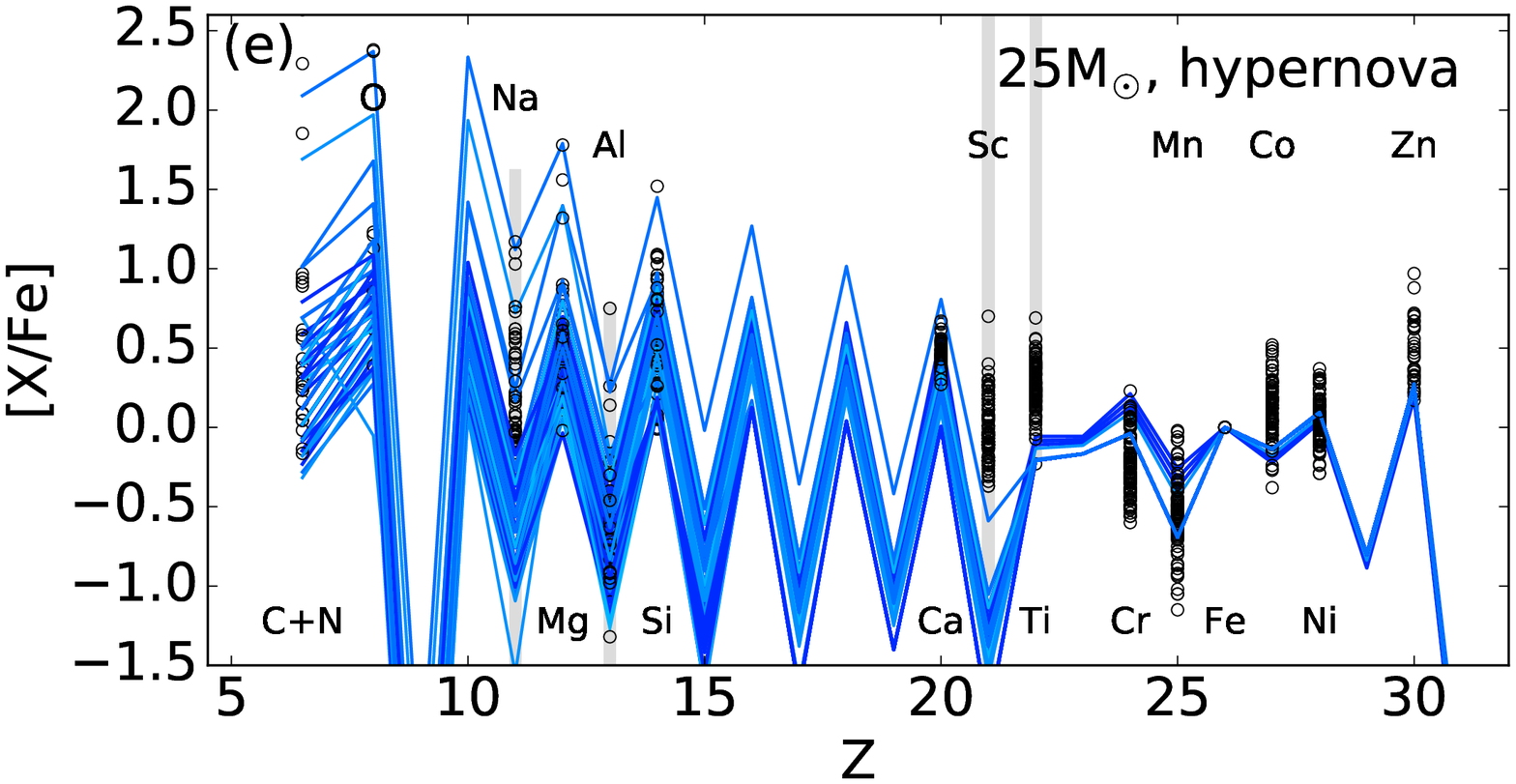}
        \end{center}
      \end{minipage}
      \begin{minipage}{0.5\hsize}
        \begin{center}
          \includegraphics[width=7.7cm]{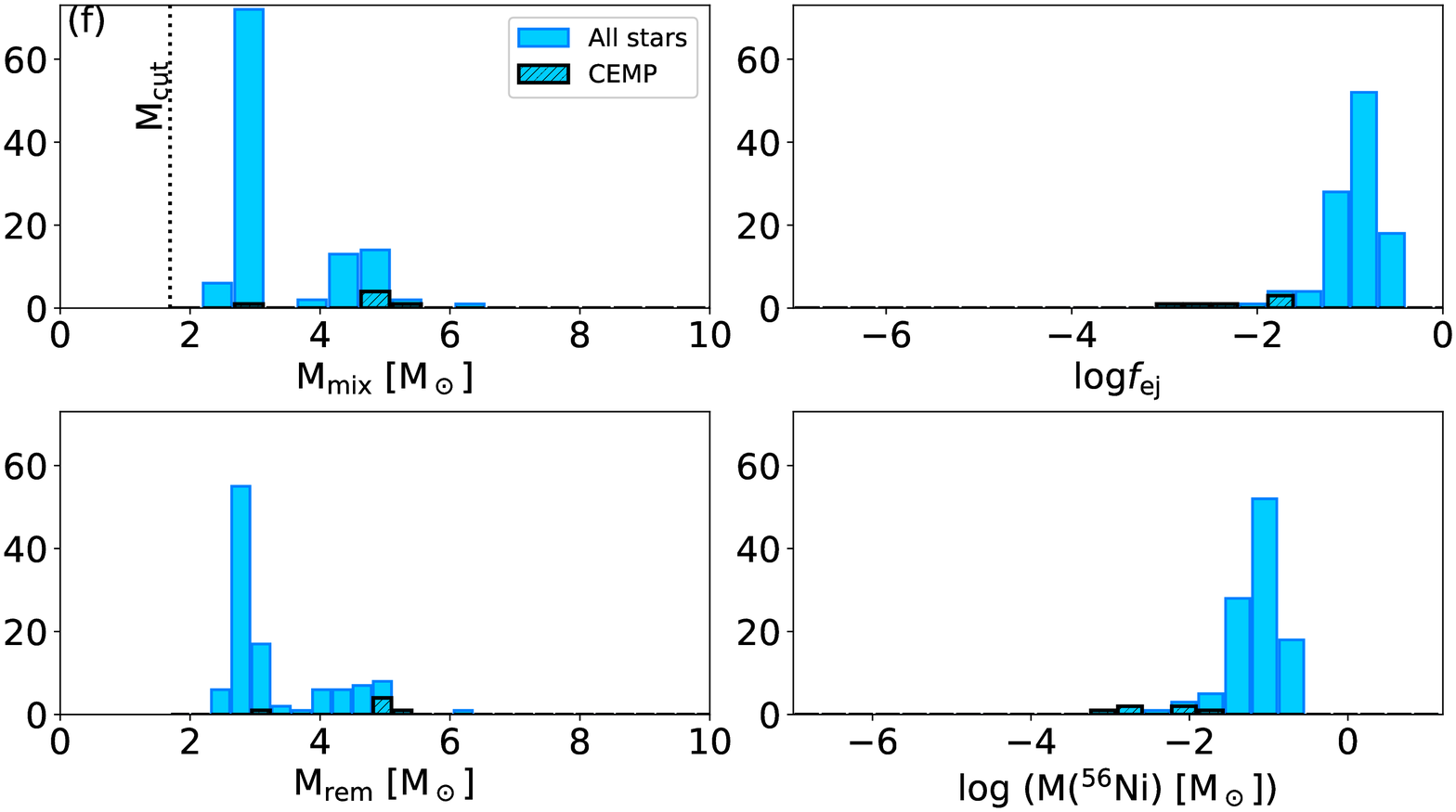}
        \end{center}
      \end{minipage}\\
    \end{tabular}
  \end{center}
  \caption{{\it Left column}: Abundance patterns of the stars best-fitted with (from top to bottom) the 15SN, 25SN, and 25HN models. The black circles indicate the observational data. The solid lines show the best-fit models where darker colors represents smaller $\chi^2_{\nu}$. {\it Right column}: The top two panels show distributions of the best-fit $M_{\rm mix}$ and $\log f_{\rm ej}$ parameters of the models shown
    in the left column. For the $M_{\rm mix}$
    parameter, the vertical lines indicate the value of
    $M_{\rm cut}$.
    The bottom two panels show distributions
    of the resulting mass of the compact remnant ($M_{\rm rem}$; Equation \ref{eq:mrem}) and the ejected
    $^{56}$Ni mass. The hatched histogram is for the CEMP stars\label{fig:abupattern-MF}.}
\end{figure*}

\begin{figure*}
  \begin{center}
    \begin{tabular}{c}
    \begin{minipage}{0.5\hsize}
        \begin{center}
          \includegraphics[width=9.0cm]{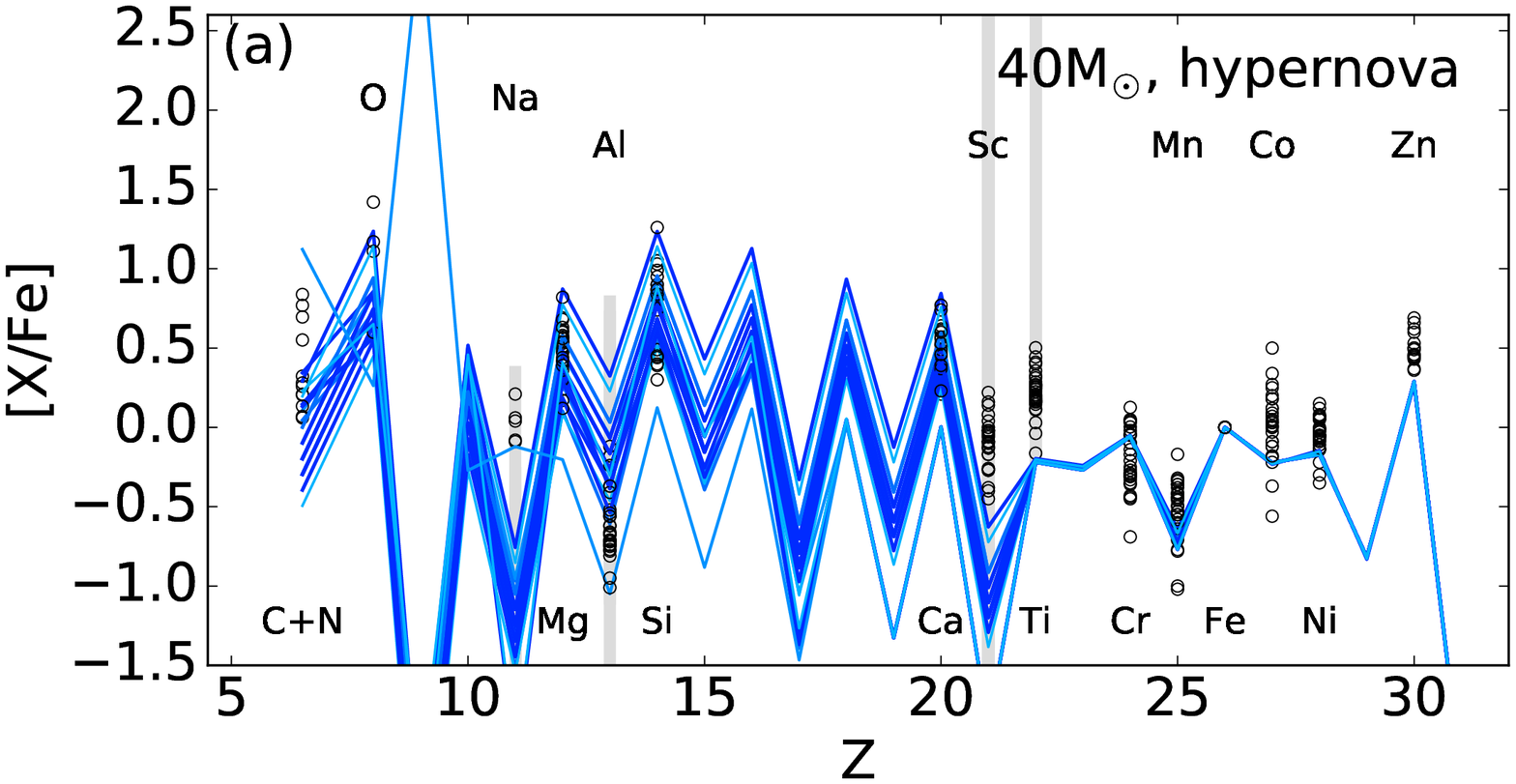}
        \end{center}
      \end{minipage}
      \begin{minipage}{0.5\hsize}
        \begin{center}
          \includegraphics[width=7.7cm]{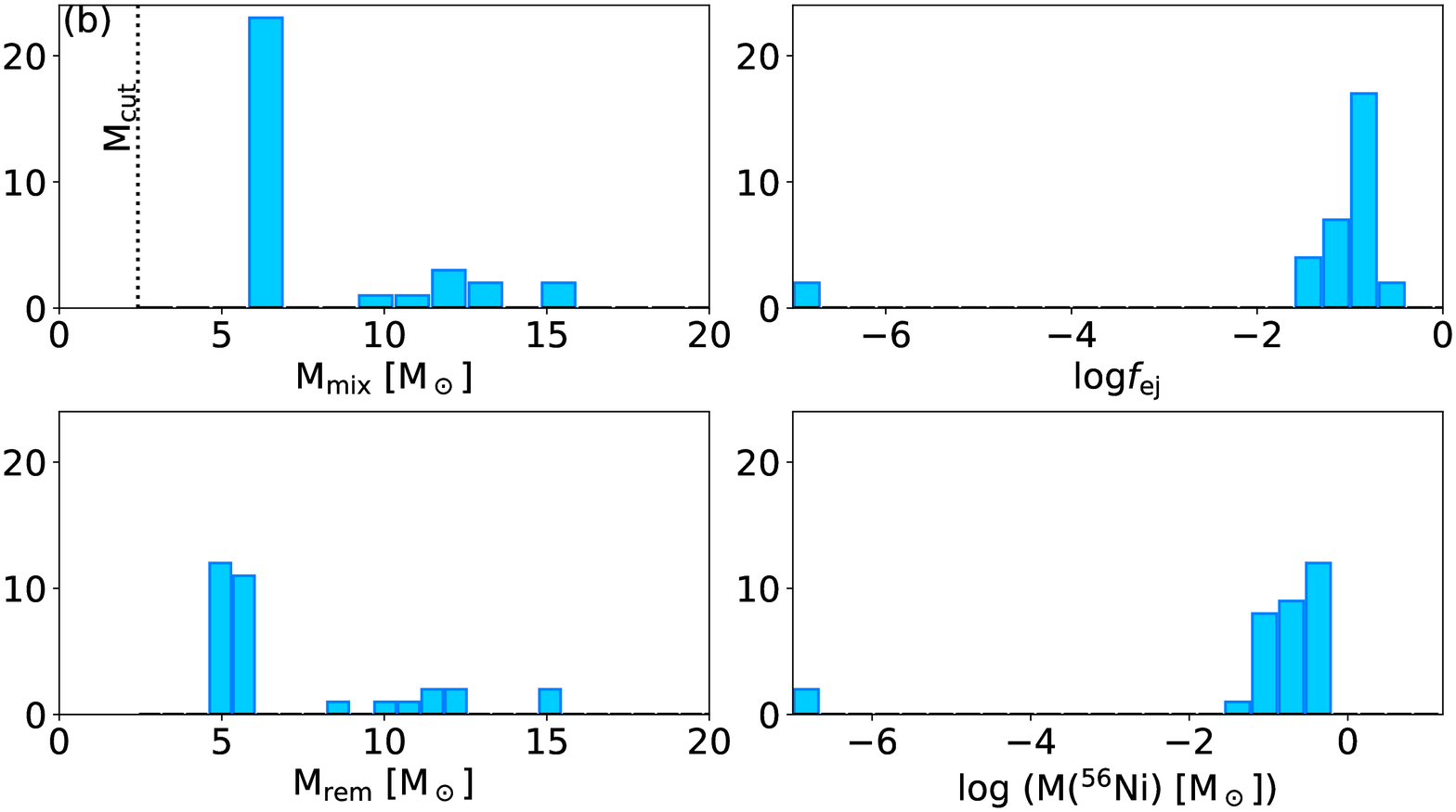}
        \end{center}
      \end{minipage}\\
      \begin{minipage}{0.5\hsize}
        \begin{center}
          \includegraphics[width=9.0cm]{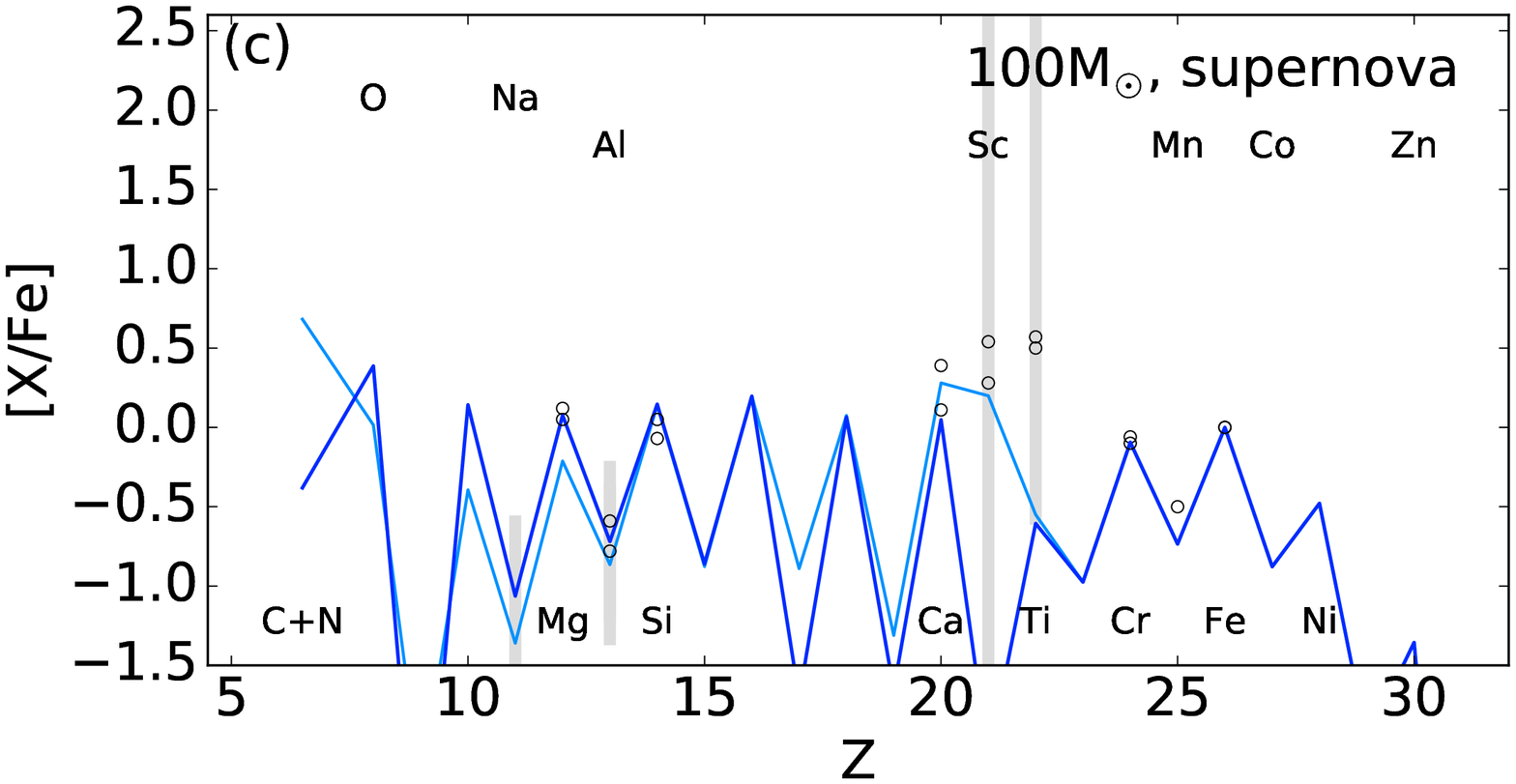}
        \end{center}
      \end{minipage}
      \begin{minipage}{0.5\hsize}
        \begin{center}
          \includegraphics[width=7.7cm]{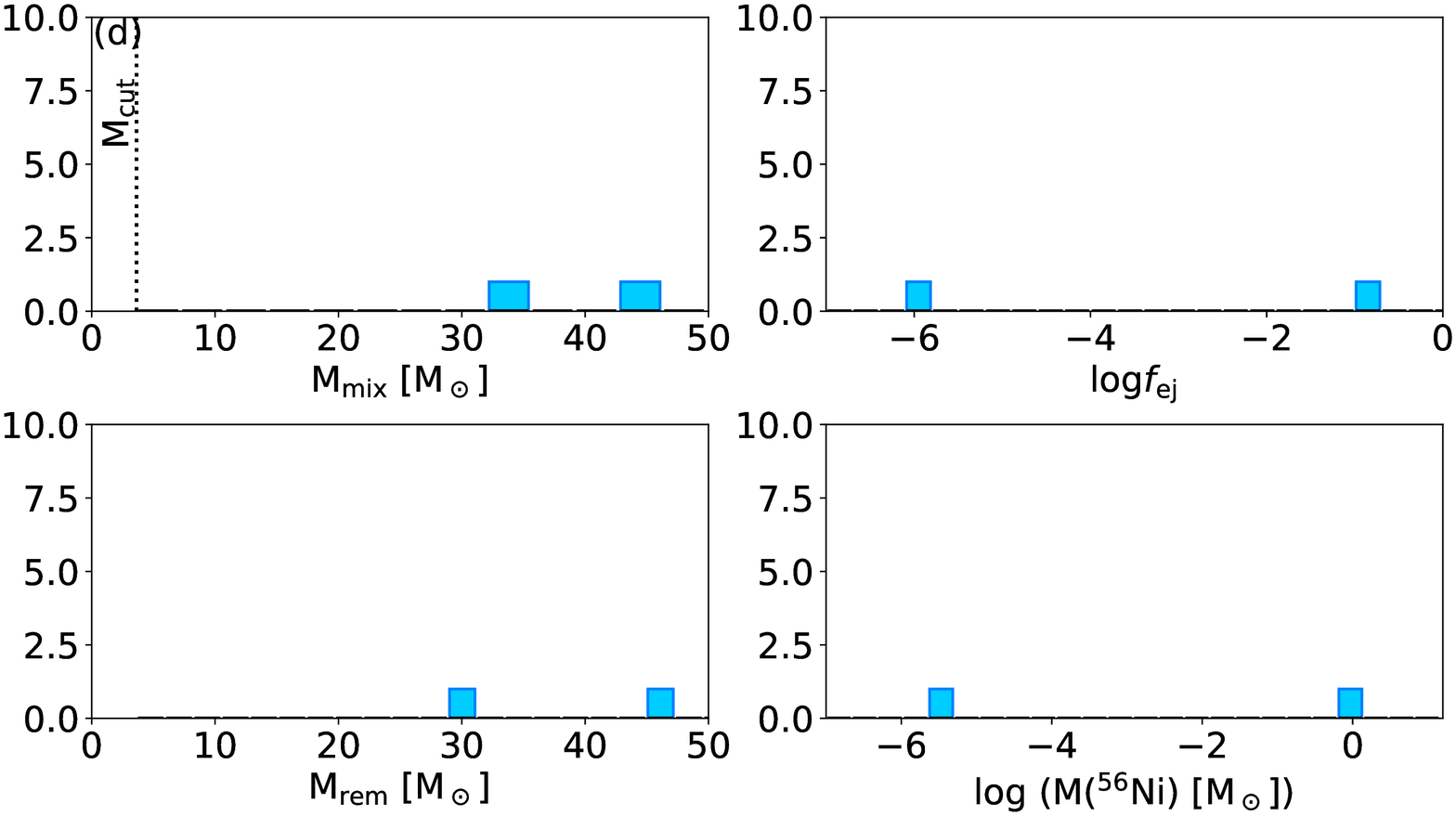}
        \end{center}
      \end{minipage}\\
    \end{tabular}
  \end{center}
  \caption{Same as Figure \ref{fig:abupattern-MF} but for the 40HN and 100SN models. \label{fig:abupattern-MF2}}
\end{figure*}

The left panels of Figures \ref{fig:abupattern-MF} and \ref{fig:abupattern-MF2}
show observed abundances of stars (circles) and the best-fit models (solid and dotted lines) for the cases of $\chi^2_{\nu}<5$. 
From top to bottom in the two Figures, the stars that are best-fitted by the
15SN, 25SN, 25HN, 40HN, and 100SN models are shown.
The right panels show histograms for the best-fit 
$M_{\rm mix}$ and $f_{\rm ej}$ values 
and the resultant remnant 
masses ($M_{\rm rem}$) and ejected $^{56}$Ni masses for each model.

\subsubsection{Stars fitted with the 15SN model \label{sec:15SN}}

Figure \ref{fig:abupattern-MF}(a) shows the observed abundances
and the best-fit $M=15 M_{\odot}$/supernova (15SN) models.
As mentioned before, 
the best-fit 15SN models are characterized by
relatively small differences in abundance ratios
between neighboring odd and even atomic-number elements among Na to Si,
compared to the other mass/energy models.
In particular, the difference between [Na/Fe] and [Mg/Fe] ratios
is small compared to the other models. 

As illustrated in Figure \ref{fig:abupattern-MF}(b), the
$M_{\rm mix}$ parameter is peaked at $\sim 1.6 M_{\odot}$, which
approximately corresponds to the outer boundary of Si layer
in the post-supernova structure. 
The $f_{\rm ej}$ parameter is
peaked at $\log f_{\rm ej}\sim -0.5$, which indicates that
$\sim 30$ \% of the mass contained in the mixing zone
($M_{\rm cut}-M_{\rm mix}$) is finally ejected and left of the
mass fallback to the compact remnant. 
As a result, as shown in the
bottom two panels of Figure \ref{fig:abupattern-MF}(b), the
mass of the compact remnant is predominantly $\sim 1.5 M_{\odot}$, which
corresponds to a neutron star,
and the ejected mass of $^{56}$Ni is in the range of $0.01-0.1 M_\odot$.

\subsubsection{Stars fitted with the 25SN model}
Figure \ref{fig:abupattern-MF}(c) shows the observed
abundances and the best-fit $M=25M_{\odot}$/supernova
models (25SN). 
Compared to the 15SN model, the [Na/Fe]
ratios are lower and predominantly subsolar. Also, on average,
the [Ni/Fe] ratios are higher for the stars fitted with the 25SN model
compared to those fitted with the 15SN model.

Figure \ref{fig:abupattern-MF}(d) demonstrates that the
large fraction of these stars are fitted with
$M_{\rm mix}\sim 2-4 M_{\odot}$, which corresponds to the
inner boundary of Si layer up to the inner part of the CO core in the
post-supernova structure. 
The ejected fraction is $0.01-0.5$, which 
results in compact remnants of $2-4 M_{\odot}$, which generally corresponds to
a black hole. The ejected
$^{56}$Ni mass is $0.01-0.1 M_{\odot}$, 
similar to those seen in the 15SN model.

\subsubsection{Stars fitted with the 25HN model}

About half of the sample stars are best fitted with the $M=25M_{\odot}$/hypernova
models (25HN), that are shown in Figure \ref{fig:abupattern-MF}(e).
On average, the [Si/Fe], [Co/Fe], and [Zn/Fe] ratios
are larger for the best-fit 25HN models than for the best-fit
25SN models. 

Figure \ref{fig:abupattern-MF}(f) shows that the large fraction of
stars are fitted with $M_{\rm mix}$
$\sim 3 M_{\odot}$, which approximately corresponds to the
outer boundary of the Si-burning layers in the post-supernova structure. 
The ejected fraction is $0.01-0.5$, which results in the remnant masses of
$2-4 M_{\odot}$
and the ejected Ni mass of $0.01-0.1 M_{\odot}$. 

Compared to the 25SN models, 
 the 25HN models have more extended regions for the explosive burning and thus 
result in larger productions of Si. Consequently, with the 
similar remnant masses, the ejected Si abundances are larger for the
25HN model. Also, the explosive Si burning produces larger amounts
of Co and Zn in the more energetic explosions,
which better fit the observed high [Co/Fe] and/or [Zn/Fe] ratios.

\subsubsection{Stars fitted with the 40HN model}

Figure \ref{fig:abupattern-MF2}(a) shows the observed
abundances and the best-fit $M_{\odot}=40M_{\odot}$/hypernova models (40HN).
In contrast to the stars best-fitted with the 25HN model, most of the
stars do not have Na measurements, which allows the best-fit models with
a very low ($<-0.5$) [Na/Fe] ratio.

Figure \ref{fig:abupattern-MF2}(b) shows that, in most cases,
the $M_{\rm mix}$ parameters are below $\sim 6 M_{\odot}$,
which corresponds to the outer boundary of the Si layer.
The ejection fractions are peaked at $\log f_{\rm ej}=0.1$, resulting in
the remnant mass of $\sim 6 M_{\odot}$, which indicates the
formation of a black hole after the hypernova.
The ejected $^{56}$Ni mass is $\sim 0.1 M_{\odot}$, which is
broadly in agreement with 
those estimated for nearby hypernovae of 
stars with the main-sequence mass $\sim 40 M_{\odot}$ \citep{2006NuPhA.777..424N}.

\subsubsection{Stars fitted with the 100SN model}

Only two stars in our sample (HE 0130-2320 and HE 0218-2738)
are best-fitted with the 100SN models, one of which
has $\chi^2_\nu<3.0$. The observed abundances and the best-fit models
are shown in Figure \ref{fig:abupattern-MF2}(c).
The abundances of these stars are characterized by [Mg/Fe]
and [Si/Fe] ratios of $\sim 0.0\pm 0.1$, which are lower 
than typical metal-poor stars.
We should note that the  Ni and Zn abundance
measurements are not available for these stars, which results
in the best-fit models with very low  [Ni/Fe] and [Zn/Fe]
ratios ($\sim -0.5$ dex).
Thus, in order to confirm the characteristic abundance patterns expected from
the 100SN model, additional abundance measurements for Ni
and Zn are crucial. In fact, if the 100SN model is
  dropped from the grid of yields, as this model is not
  theoretically motivated (see Section 2.1), the two stars
  are alternatively best-fitted with the 25HN models, which
  predict [Zn/Fe]$\sim 0.3$ dex.

As shown in Figure \ref{fig:abupattern-MF2} (d), the best-fit
parameters for the two stars are $M_{\rm mix}=$46 and 34$M_{\odot}$ and
$f_{\rm ej}=10^{-6}$ and $0.16$, respectively. 
The corresponding remnant masses are $46$ and 
$29 M_{\odot}$, respectively, which suggests the formation of
a black hole with these masses.

\subsubsection{Other models}
\begin{figure}
  \plotone{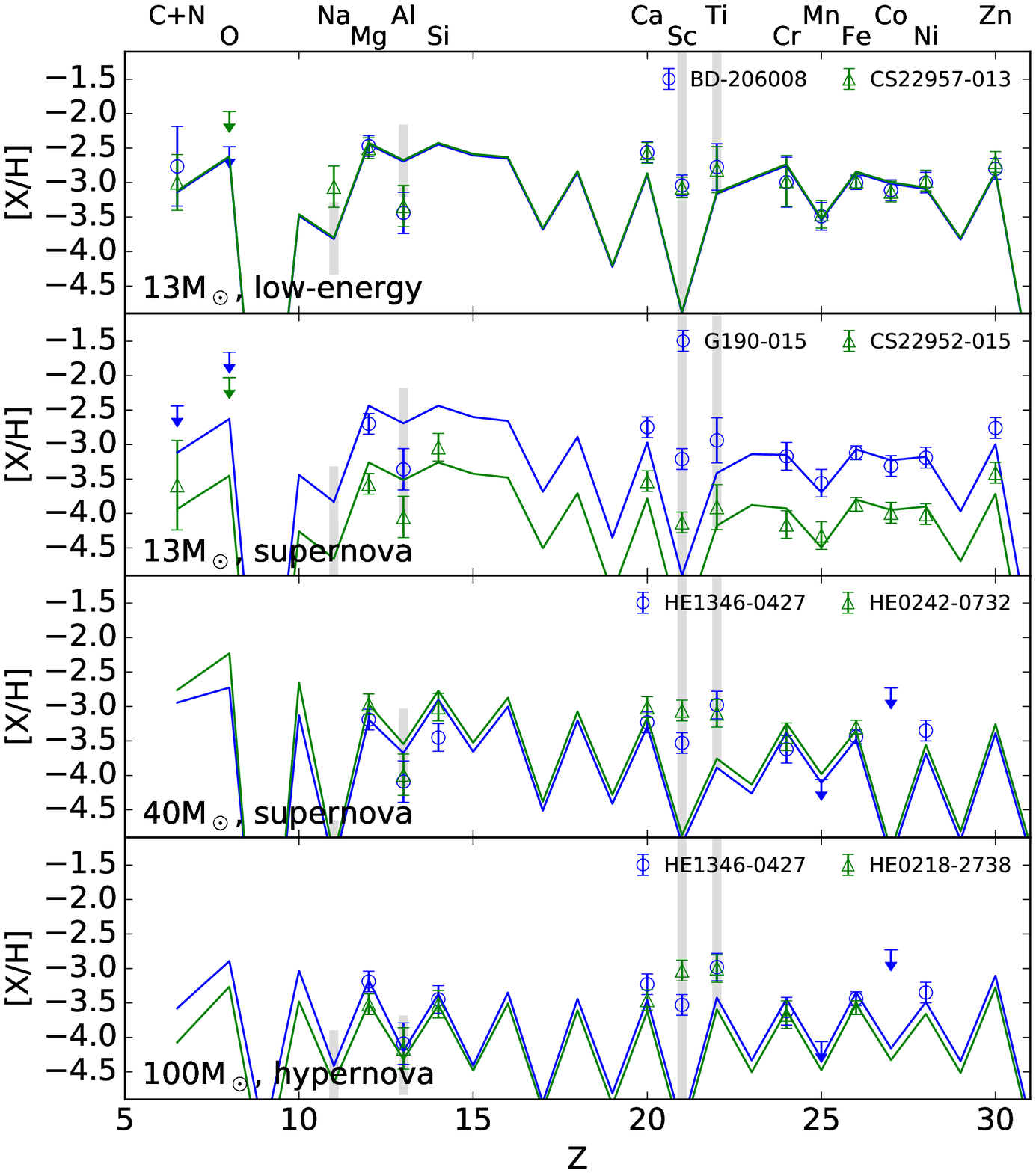}
  \caption{Abundance patterns of the 13LE, 13SN, 40SN, or 100HN models, which
    do not best-fit the data for any of the present sample stars.
    The data are shown by symbols with
    error bars and the models with relatively small $\chi^2_{\nu}$ are shown
    by solid lines with colors corresponding to the data.\label{fig:badfit}}
  \end{figure}

In our sample, stars that are best-fitted with
the 13SN or 40SN models are not found and the two objects best-fitted
with the 13LE or 100HN models but with $\chi^{2}_{\nu}>3$.
Figure \ref{fig:badfit} shows example for the fitting of these
models.
It can be seen that the 13SN model (top panel) under-produces the [Na/Fe]
ratios while they overproduce [Al/Fe]. 
The 40SN models also predict higher [Al/Fe] than the observed
values.

\subsubsection{CEMP}

\begin{figure*}
\plotone{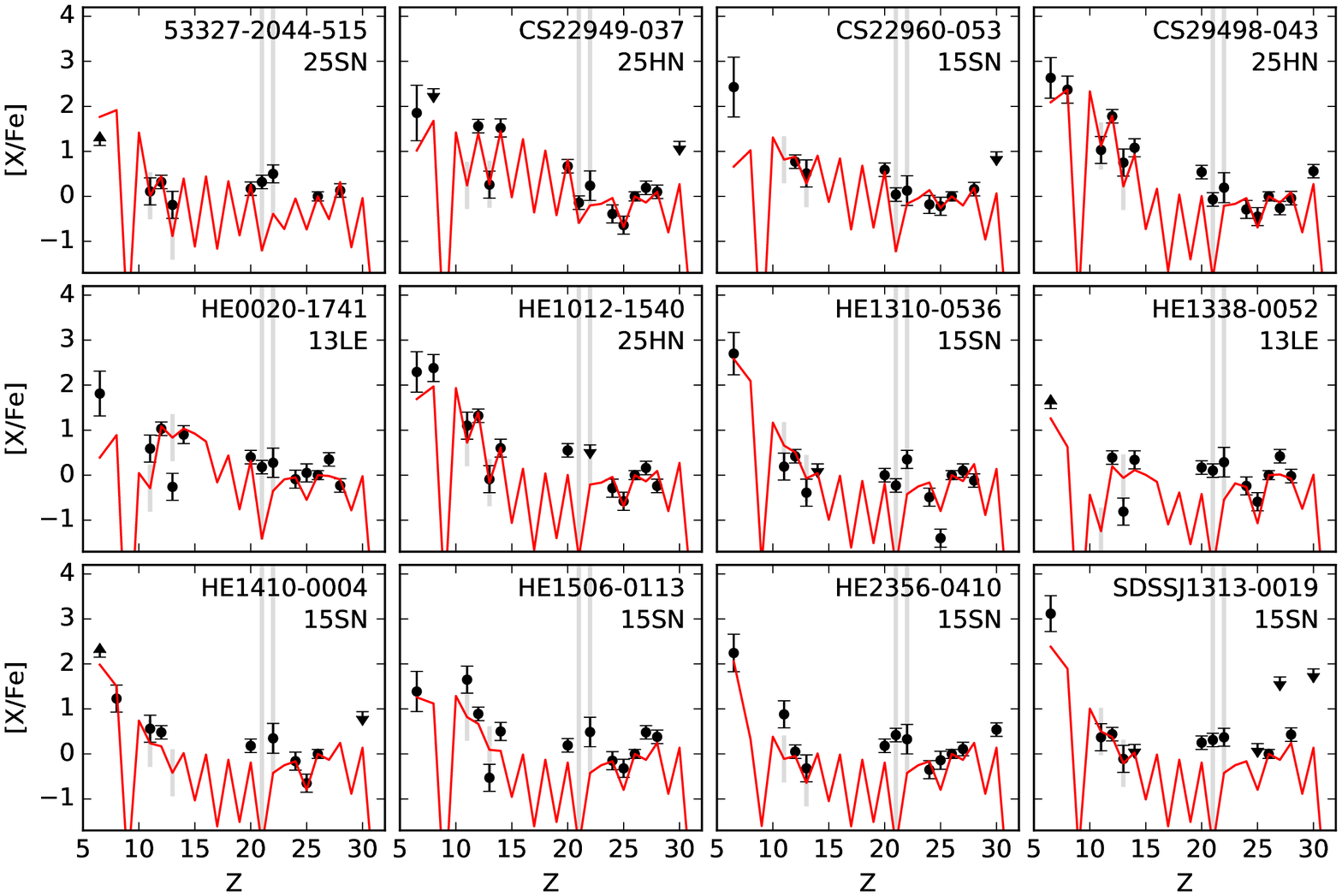}
\caption{The best-fit models for the CEMP stars with $\chi^{2}_{\nu}<5$.\label{fig:cemp}}
  \end{figure*}

In our sample, 18 stars are CEMP stars with [(C+N)/Fe]$>1.0$.
Similar to the other stars, the CEMP stars are predominantly best-fitted with
either the 15SN, 25SN, or 25HN models.
Figure \ref{fig:cemp} shows the best-fit models for the 12 CEMP stars
with $\chi^{2}_{\nu}<5$.
 Their abundances
 require the models with a larger scale mixing and fallback
 than those required for the other EMP stars; 
 the $M_{\rm mix}$ much larger than the outer boundary of the
 Si layers and the $f_{\rm ej}<0.1$. 
Consequently, the compact remnants of the CEMP progenitors span
the highest mass range in the remnant mass distribution as can be seen
in Figures \ref{fig:abupattern-MF}(b), (d), and (f).

The C$+$N enhancements in the CEMP stars are sometimes
associated with enhancements of Na, Mg, or Al abundances. In our
analysis, stars with [Mg/Fe]$>1$ (CS 22949-037,
CS 29498-043, and HE 1012-1540) 
are best-fitted by the 25HN models while those with
lower [Mg/Fe] ratios are fitted with either the 13LE, 15SN or
25SN models. 
   The requirement for high explosion energy
   stems from the fact that Mg is explosively synthesized
   at the bottom of the He layer ($M_r\sim 5.5M_\odot$).
   Consequently, in order to reproduce the high [(C+N)/Fe] ratios,
   the layer containing these explosively synthesized
   Mg should be ejected in the model, which 
   explains both the high [(C+N)/Fe] and [Mg/Fe] ratios.

\subsection{Pop III masses and explosion energy \label{sec:PopIIImass}}

As mentioned earlier, the observed abundances of the EMP sample stars 
are best explained by the models for Pop III stars with masses
$M=$15, 25, 40 or 100$M_{\odot}$ that explode with normal ($E_{51}=1$) or higher
explosion energies ($E_{51}>1$). In this section, we examine the typical
masses of the first stars whose nucleosynthetic products are incorporated
into the EMP stars.

\begin{figure*}
\plottwo{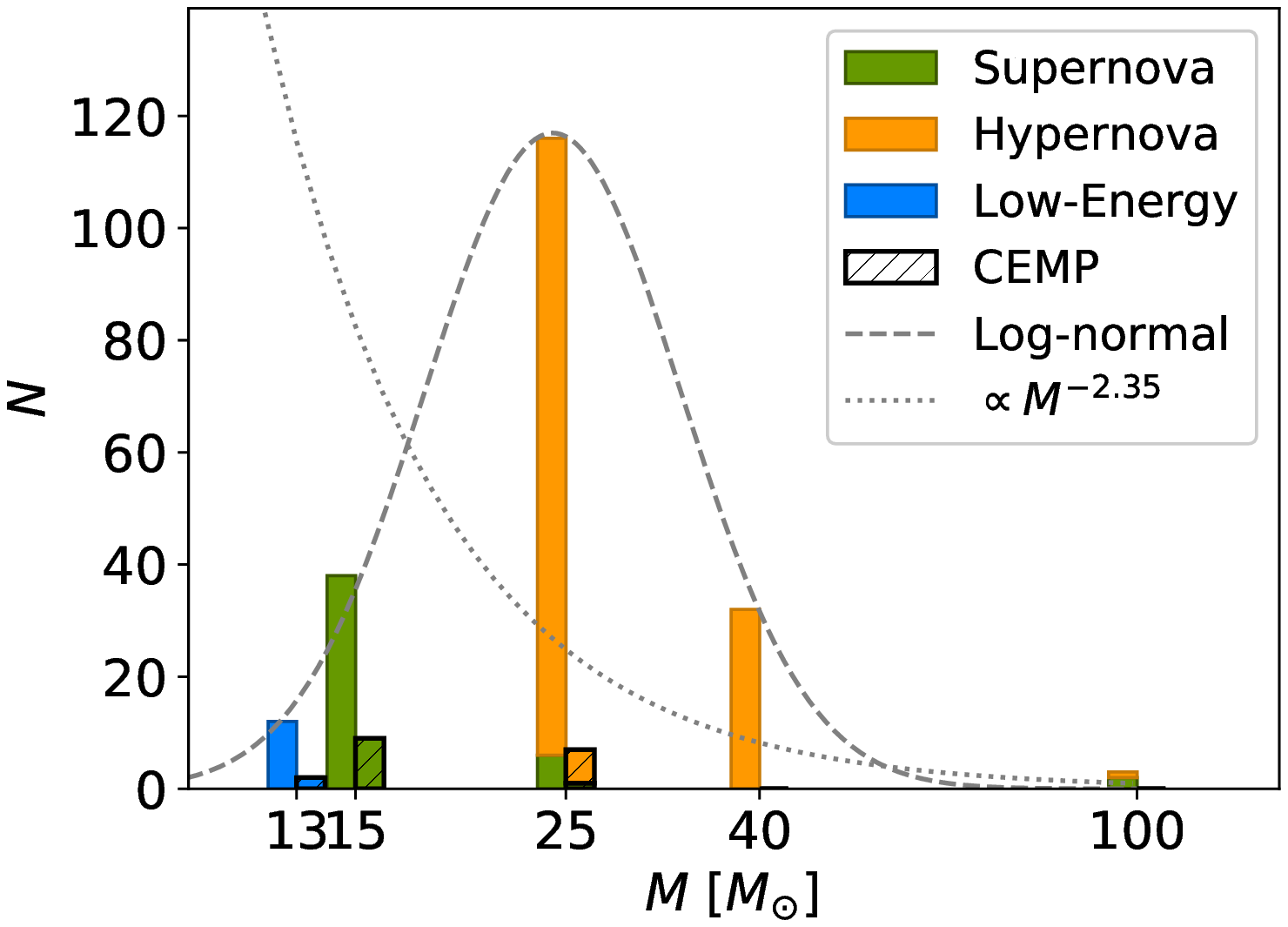}{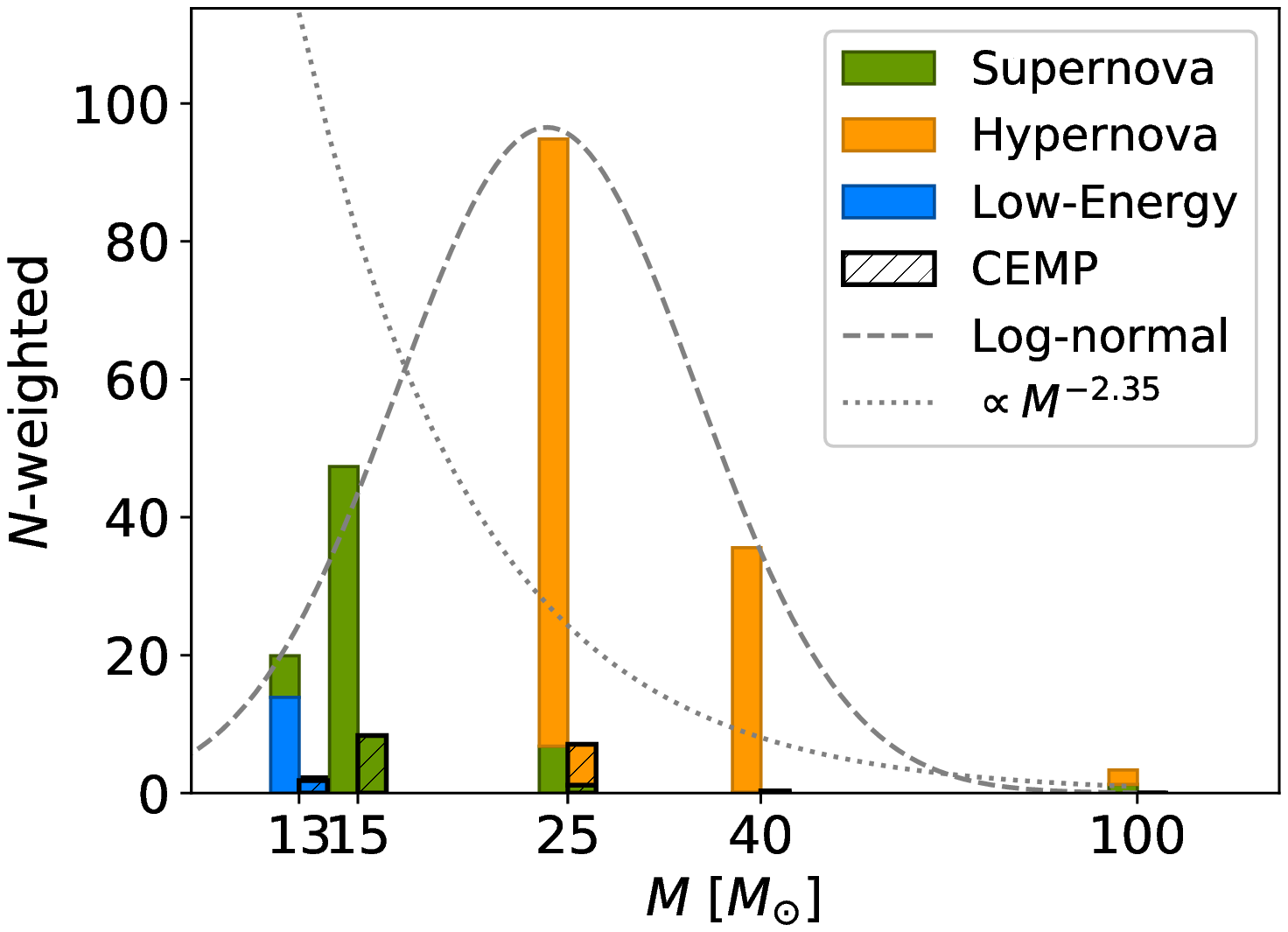}
\caption{
  {\it Left}: The histogram of the Pop III progenitor masses of the
  best-fit models. Green, orange and blue bars correspond to
  supernova, hypernova, and low-energy explosion models, respectively. 
  The results for the CEMP star are shown by
  hatched bars. The gray dashed line shows the best-fit
  log-normal function to the histogram.
  For a reference, the Salpeter IMF ($\propto M^{-3.5}$), arbitrary normalized
  for better visibility, is shown by a dotted line.
  {\it Right}: The histogram obtained by
  counting contributions from all 9 mass-energy models weighted by
  the p-value (see text for the definition).\label{fig:mhist_SNHN}}
\end{figure*}

The left panel of Figure \ref{fig:mhist_SNHN} shows the histogram of
the Pop III masses of the best-fit models. Blue, green and orange
bars correspond to the low-energy, normal-energy and hypernovae explosion
models, respectively. 
In order to include the contributions from models other than
the best-fit ones, we show in
the right panel of Figure \ref{fig:mhist_SNHN} 
the histogram obtained
by counting contributions from all 9 models weighted by
the $p$-values as, $Cp$, where the constant $C$ is set so that these weights
sum to unity for each star.

It can be seen from both histograms that the highest contribution comes from the 
$M=25M_{\odot}$ models and more than half of the whole stars are best explained by the $M=25 M_\odot$/hypernova model. 
The Pop III masses for the progenitors of the CEMP stars shown by the
hatched histograms also dominate
at $M\leq 25 M_\odot$ while relative contribution from $M=15M_{\odot}$
Pop III models are larger than the $M=25M_\odot$ models.
Given the theoretical and observational uncertainties, we found
to be difficult to clearly distinguish the $M=15$ and $25 M_{\odot}$ models. 
Therefore, the Pop III progenitors of the CEMP is not
clearly distinguishable in terms of masses,
from those of the majority of the EMP stars. 
This result implies that the physical mechanism of the formation of
CEMP stars is rather related to the state of
the mixing and fallback (e.g., Figure \ref{fig:abupattern-MF}), which presumably
occur in aspherical supernova/hypernova explosions.

\subsection{Robustness of the best-fit progenitor masses\label{sec:testerr}}

\subsubsection{Effects of observational uncertainties\label{sec:obserr}}
We examine the robustness of the fitting results against
the fiducial observational errors assigned to the data
(0.1-0.3 dex; see Section \ref{sec:obsdata}).
For this test, we select objects with
$\chi^2_{\nu}<1$ among those best-fitted with each of the
15SN, 25SN, 25HN, 40HN, 100SN and 100HN models. 
For each of the selected objects, the same abundance
fitting procedure is performed 
100 times by adding noises taken from a Gaussian distribution
with a sigma (standard deviation) equals to the adopted observational errors.

\begin{figure}
  \plotone{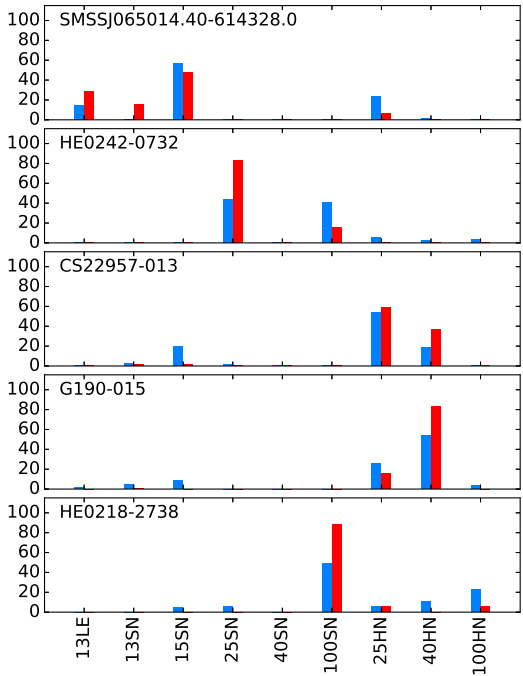}
  \caption{Distributions of the best-fit models obtained from the 100 abundance-fitting
    runs by adding noises from the observational errors.
    From top to bottom, the results for
    objects originally best-fitted with the 15SN, 25SN, 25HN, 40HN and 100SN models
    (SMSS J065014.40-614328.0, HE 0242-0732, CS 22957-013, G 190-15, and HE 0218-2738)
    are shown. The blue histograms show the results for the
    fiducial observational errors (0.1-0.3 dex) assigned as described in Section \ref{sec:obsdata}.
    The red histograms show the results when half of the fiducial errors are assigned.
    \label{fig:mc_hist}}
  \end{figure}

The blue histogram in each panel of Figure \ref{fig:mc_hist} shows the
distribution of the best-fit progenitor
mass/energy models obtained from the 100 abundance-fitting runs. From top to bottom
panels, the results for the objects originally best-fitted with the
15SN, 25SN, 25HN, 40HN, 100SN and 100HN models
 are shown. 

For the adopted observational errors,
a different progenitor mass/energy
model is chosen as the best-fit in some cases. For the case of
SMSS J065014.40-614328.0, which is originally best-fitted with
the 15SN model, a different model, either the 13LE, 13SN, 25SN or 25HN model,
best-fits the data more than 50 times. 
The probability of getting other
models as the best-fit is also high for the object
fitted with the 25SN model (HE 0242-0732), for which there is
$\sim 40$ \% probability of obtaining the 100SN model as the best fit.
On the other hand, for the objects
best-fitted with the 25HN, 40HN or 100SN models,
the original best-fit models are chosen
with $\gtrsim 50$ \% probability.

These results suggest that, for the fiducial observational errors
adopted as in Section \ref{sec:obsdata}, 
the best-fit models are not always robustly determined, especially
for objects best fitted with the 25SN model.
On the other hand, 
the objects best-fitted with the
25HN, 40HN, or 100SN models tend to have distinct abundance patterns, so that
they are more robustly distinguished.

The recovery of the original fit is improved when the adopted
observational errors are half the fiducial value ($0.05-0.15$ dex), as shown by the
red histograms in Figure \ref{fig:mc_hist}. All but one object are fitted by the
original best-fit models for more than 50\% of the runs.
Therefore, reducing the observational errors is crucial to obtain
tighter constraints on the progenitor mass distributions of
Pop III stars in our analysis.  

\subsubsection{Effects of systematic uncertainties}

\begin{figure*}
  \plottwo{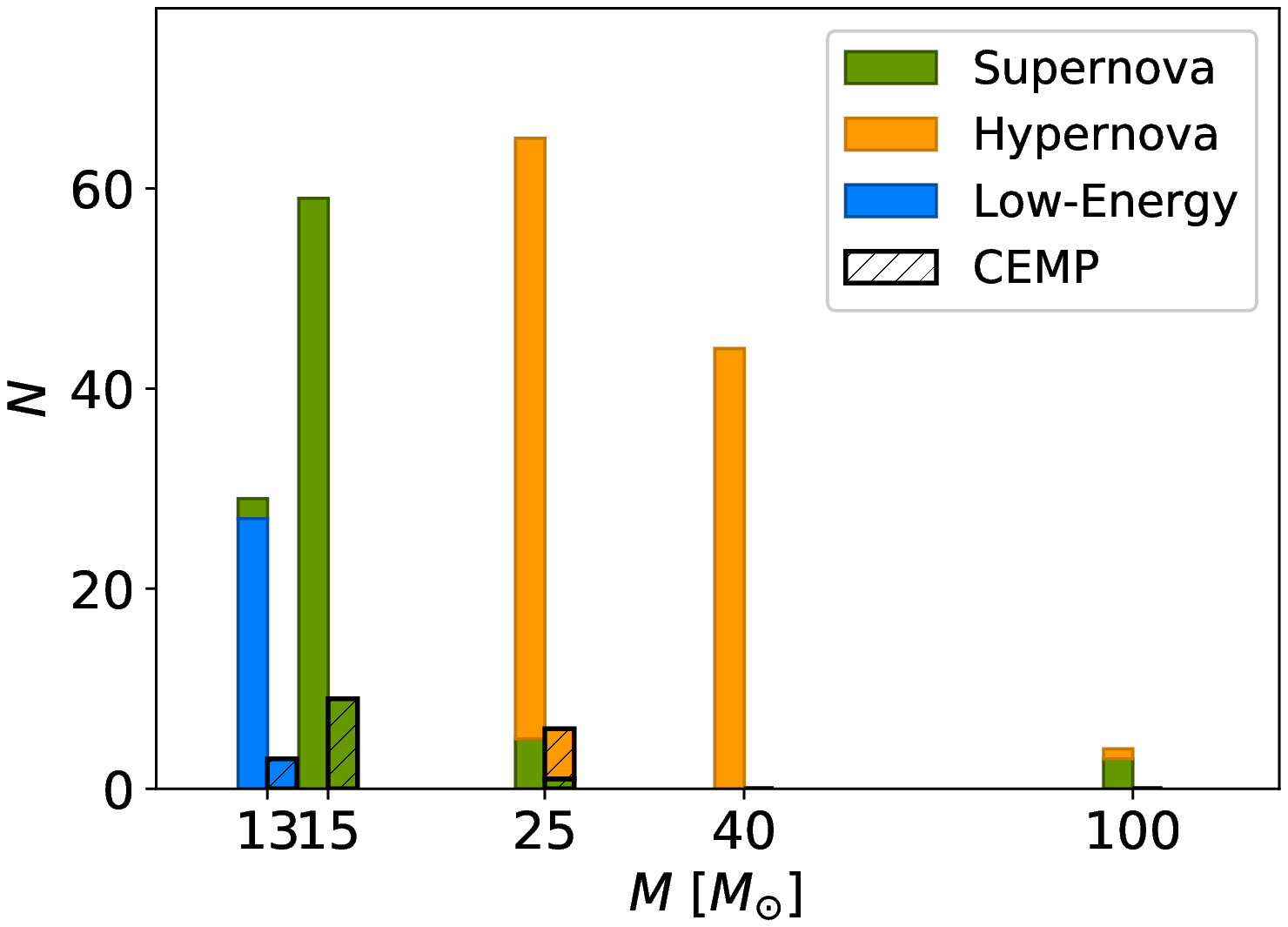}{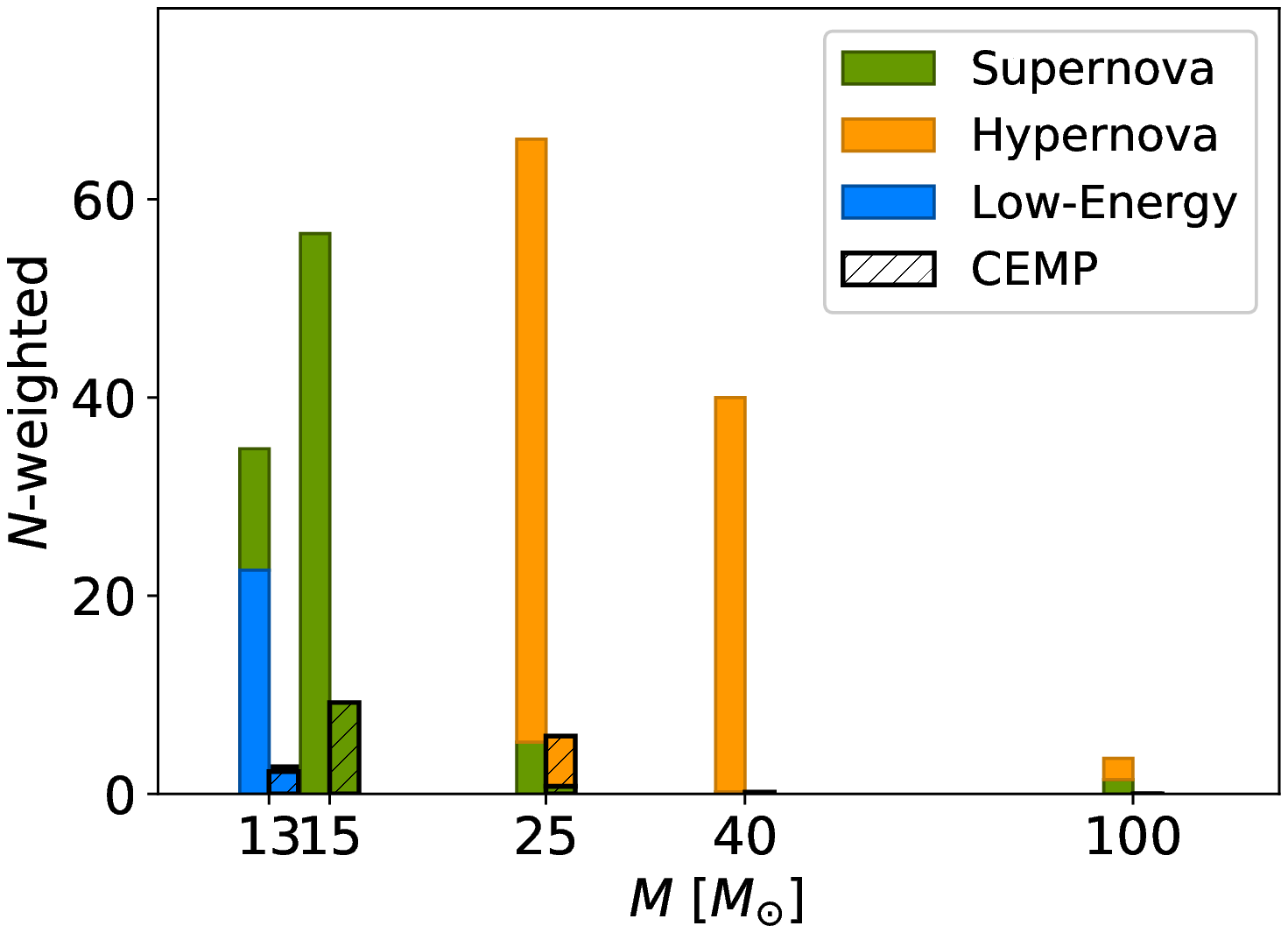}
  \plottwo{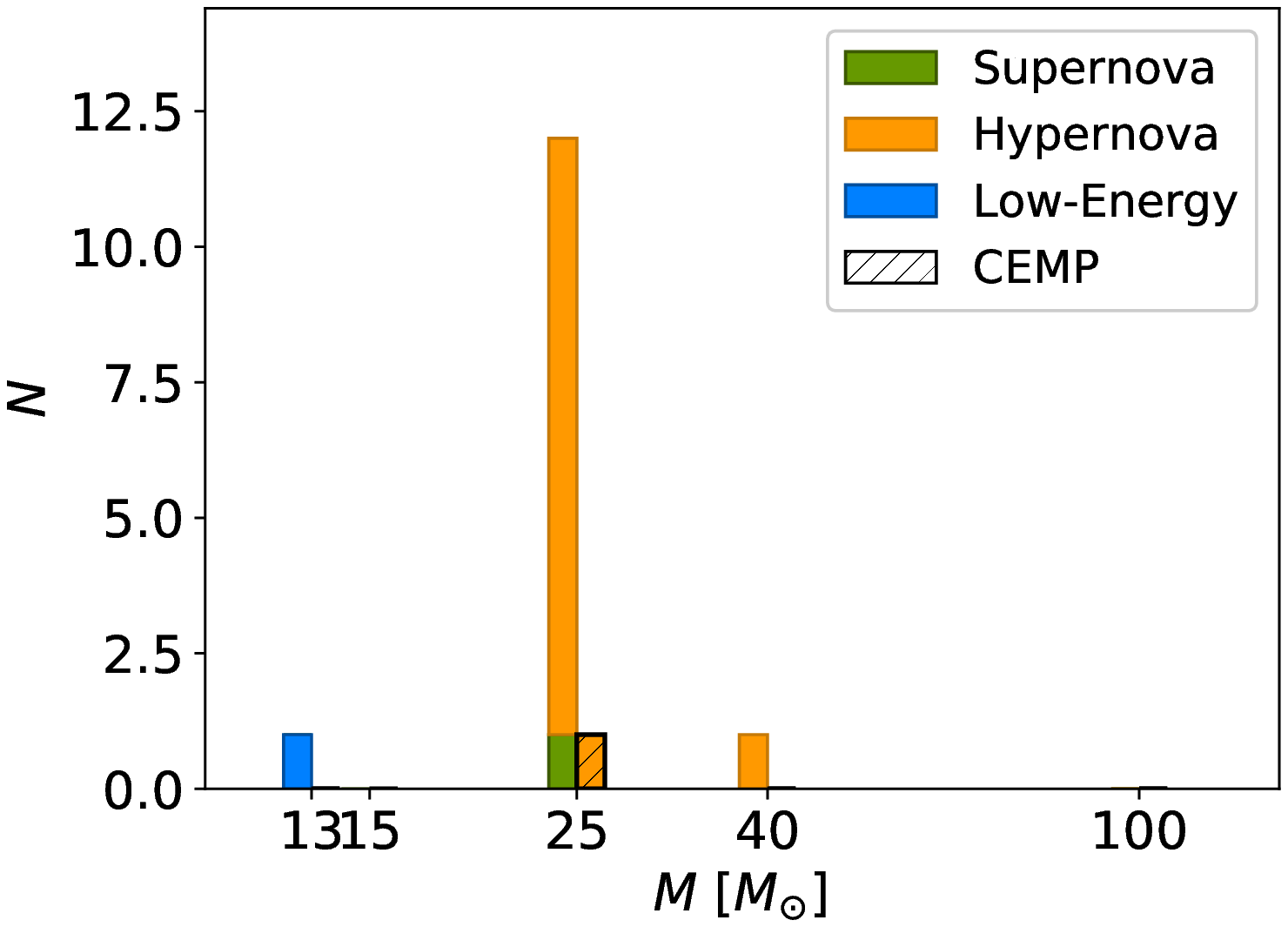}{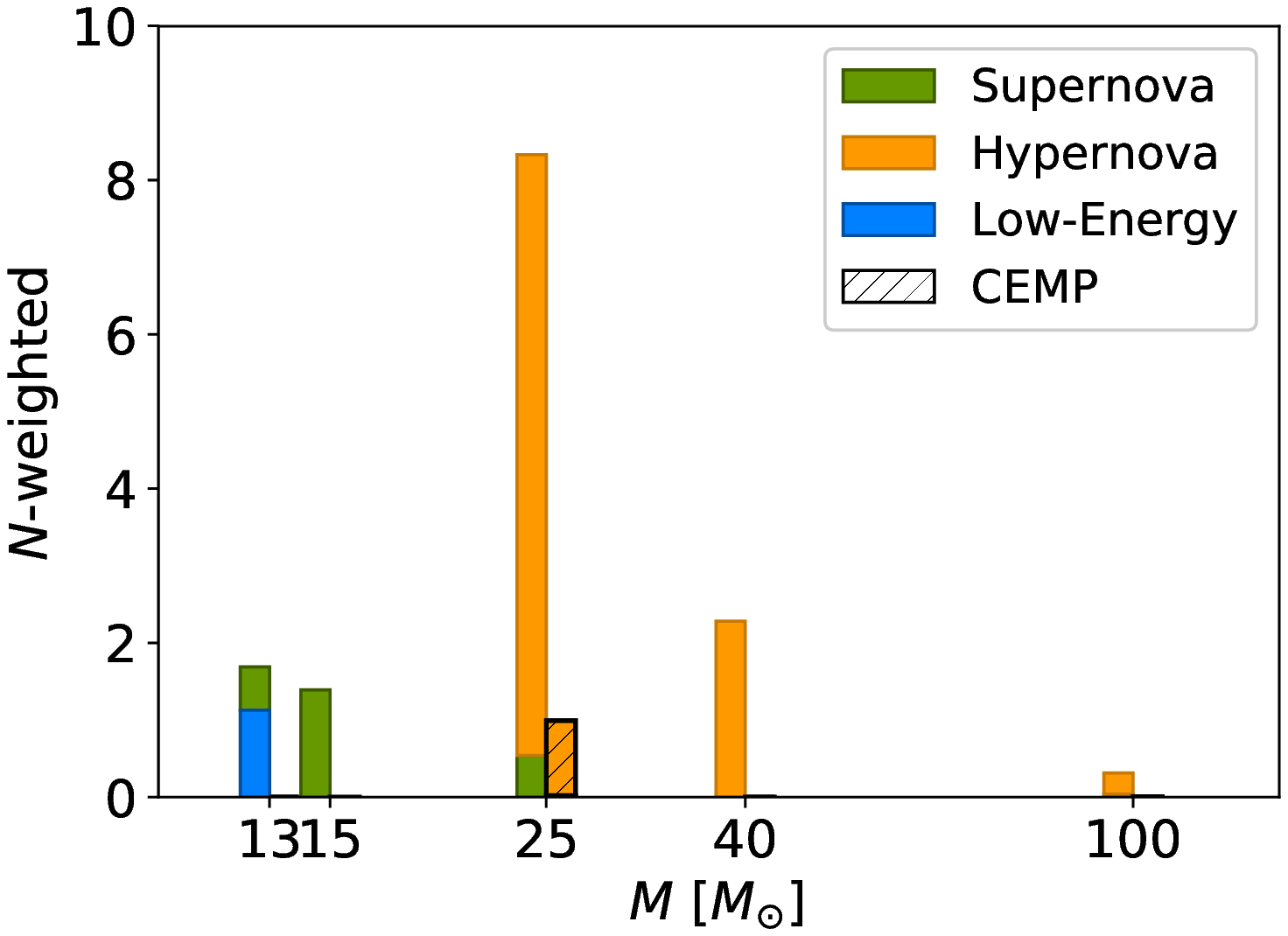}
  \caption{{\it Top}: Same as Figure \ref{fig:mhist_SNHN} but for the result of the abundance fitting
    after the uniform NLTE correction to the observed Al abundance is applied.
    {\it Bottom}: Same as Figure \ref{fig:mhist_SNHN} but
    only for objects for which abundance measurements or upper limits of
    all the elements we chose are available. \label{fig:mhist_systematics}}
 \end{figure*}

One of the major systematic uncertainties in measured abundances
comes from the NLTE effect on Al abundances,  for which
suggested NLTE correction is up to $\sim +0.6$ dex.
The effect of change in measured Al
abundances is tested by repeating the same abundance fitting procedures
but adding 0.6 dex to the Al abundances measured
under the assumption of LTE as a NLTE correction.
The resultant progenitor-mass
histogram and the histogram obtained by weighting with the $p-$value  are 
shown in the top two panels in Figure \ref{fig:mhist_systematics}.
As can be seen from the top-left panel, some fraction of stars that
have originally been fitted with the 25HN model are now better fitted with
either the 13SN or 15SN models. This can be understood since the elevated
Al abundance via the NLTE correction give smaller odd-even effect
and thus better fitted with a lower progenitor mass model
as mentioned in Section \ref{sec:15SN}. Therefore, the weighted
histogram on the top-right panel shows that the contribution
from $M\le 15M_{\odot}$ is larger than that of $M=25M_{\odot}$ in contrast
to the original histogram. This results highlight the importance of
obtaining the NLTE abundances for Al to discriminate the progenitor
Pop III masses between $M\le 15$ and $25M_{\odot}$ in observations.

Another possible source of systematic uncertainties is
missing observational data for certain elements on the Pop III
masses. 
For example, as can be seen from the bottom panel of Figure \ref{fig:representative},
the objects fitted with the 100SN model tend to have
smaller number of elements measured in observations than the stars
fitted with the other models. This might imply that
the finding of the best-fit with the 100SN models could stem from
the non-measurements of particular elements. In order to test the
robustness of the Pop III mass histogram 
against the number of constraints, in the bottom two panels in
Figure \ref{fig:mhist_systematics}, we plot the same histogram as
in Figure \ref{fig:mhist_SNHN}
only for stars with the abundance constraints, either measurements or upper limits,
are available for all the elements we chose
(C, N, O, Na, Mg, Al, Si, Ca, Sc, Ti, Cr, Mn, Fe, Co, Ni, and Zn).
 This reduces the sample stars to only 15. 
 The resulting histogram for the best-fit model is shown
 on the bottom-left panel
 and the corresponding $p-$value-weighted histogram is shown
 on the bottom-right panel. The both histograms show that the
 Pop III masses are peaked at $25M_{\odot}$ and that,
 compared to the histogram for the
 whole sample (Figure \ref{fig:mhist_SNHN}), the contribution from
 the 15SN models is suppressed.
It can be seen, however, that the main results on the Pop III masses, namely,
the mass distribution is dominated by $M<40M_{\odot}$ 
and is peaked at $M=25M_{\odot}$, are robust against the number of
abundance constraints in our analysis.

To further test the non-measurement of
  a specific element in our abundance fitting, we select  
stars that have the measurements of all 16 elements we chose. 
In our sample, only one star CS29498-043 has the complete abundance
measurement. We perform the abundance fitting by excluding
an element one by one and compare the results
with the original best-fit model ($M=25 M_\odot$ and $E_{51}=10$).
With this experiment, except for the cases omitting Si or Zn,
the original best-fit model is reproduced but with slightly different
mixing-fallback parameters. 
The lack of Si measurement leads to a best-fit model
with a higher progenitor mass ($M=40 M_\odot$)  and a higher explosion
energy ($E_{51}=30$). On the other hand, the lack of Zn measurement
leads to a best-fit model with the same progenitor mass and a lower
explosion energy ($E_{51}=1$), as expected.  
This experiment demonstrates that the best-fit model does not
change even if we omit one of the abundance measurements except for
Si or Zn. Therefore, in our analysis, abundance measurements of Si and Zn are
particularly important in constraining the Pop III models.

\subsection{Comparisons with previous studies}

\citet{2015ApJ...809..136P} fits supernova yields of Pop III stars
to a sample of 20 ultra-metal-poor stars using the publically available
code, {\it STARFIT}, which is based on the grids of yields
calculated by \citet{2010ApJ...724..341H}.
Although the {\it STARFIT} also takes into account the mixing-fallback
process, the assumption in the code is different
  from ours. In the {\it STARFIT}, the supernova explosions
  are treated close to rather spherically symmetric and the Rayleigh-Taylor
  instability and spherical fallback are mainly assumed to be the mixing-fallback mechanism. On the other hand, we consider
 various degrees of asymmetry
  in the explosions including those associated with a jet and fallback along the equatorial plane.
 
The comparison for 10 stars analyzed in common is summarized in Table
\ref{tab:comp_placco}. \citet{2015ApJ...809..136P} obtained the best-fit
Pop III models ranging from $M=$10.9 to 28 $M_\odot$ with the explosion energies
$E_{51}=0.3-10.0$.
The range in the progenitor masses are broadly consistent with
the results obtained with our analysis (15 or 25$M_\odot$).

Differences in explosion energies can be seen in some of these
stars. For example, 
explosion energies are lower in \citet{2015ApJ...809..136P} than in our study
for many of the stars listed in Table \ref{tab:comp_placco}. 
In the {\it STARFIT} code applied in \citet{2015ApJ...809..136P},  
  the amount of fallback is coupled to the explosion energy
  based on 1D hydrodynamical simulations by \cite{2008ApJ...679..639Z}
  and thus the low explosion energy is required for
 the larger fallback \citep{2010ApJ...724..341H}.
For example, Pop III SNe with $E_{51}\le0.6$ 
result in larger fallback leaving behind a compact remnant
with larger masses. 
The mixing is assumed to be resulted from Rayleigh-Taylor mixing
 and is parametralized
by a fraction of the He core mass, $f_{\rm mix}$ corresponding to
a width of a box-car smoothing kernel for the abundance structure.
The difference is also
partly due to the limited progenitor mass and energy coverage
while allowing a wider range of the mixing and fallback
parameters in the present study.

Table \ref{tab:comp_placco} also lists inferred mass of the
compact remnant and the ejected mass of $^{56}$Ni from
\citet{2008ApJ...679..639Z} and \citet{2010ApJ...724..341H}. 
The remnant masses are systematically higher in \citet{2015ApJ...809..136P}
ranging from $\sim$1.5 $M_{\odot}$ for $M=10-15$ $M_{\odot}$
progenitor models and $\sim 7.9-12.0$ $M_{\odot}$ for $M>20M_{\odot}$
progenitor models.  
On the other hand, the remnant masses
in our study range from $M=$1.5 to 5.0$M_{\odot}$.
The ejected $^{56}$Ni masses are smaller with
$M(^{56}$Ni$)<10^{-3}M_{\odot}$
in \citet{2015ApJ...809..136P}, while they are
typically $M(^{56}$Ni$)=10^{-4}-10^{-1} M_{\odot}$ in the present analysis.

\begin{deluxetable*}{lcccccccccc}
  \tablecaption{Comparison with \citet{2015ApJ...809..136P} \label{tab:comp_placco}}
\tablehead{
  \colhead{Star name} & \colhead{[Fe/H]} & \colhead{[C/Fe]}& \multicolumn{4}{c}{Placco+15}& \multicolumn{4}{c}{This work}  \\
   & &  &\colhead{$M$} & \colhead{$E$} & \colhead{$M_{\rm rem}$\tablenotemark{a}} & \colhead{$M(^{56}{\rm Ni})$\tablenotemark{b}} &
  \colhead{$M$} & \colhead{$E$} & \colhead{$M_{\rm rem}$} & \colhead{$M(^{56}{\rm Ni})$}\\
     & & & \colhead{($M_\odot$)} & \colhead{($10^{51}$ erg)} & \colhead{($M_\odot$)} & \colhead{($M_{\odot}$)} &
 \colhead{($M_\odot$)} & \colhead{($10^{51}$ erg)} & \colhead{($M_\odot$)} & \colhead{($M_{\odot}$)}
  }
\startdata
\multicolumn{11}{c}{[C/Fe]$<1$}\\
CS 30336-049 & $-4.03$ & $+0.09$  & 21.5 & 0.3 & 7.88 & 7.66$\times 10^{-5}$ & 15 & 1 & 1.49 & 6.9$\times 10^{-2}$ \\
HE 1424-0241 & $-4.05$ & $+0.63$  & 21.5 & 0.3 & 7.88 & 7.66$\times 10^{-5}$ & 15 & 1 & 2.01 & 3.4$\times 10^{-2}$ \\
CD-38 245 & $-4.15$ & $-0.09$  & 21.5 & 0.3 & 7.88 & 7.66$\times 10^{-5}$ & 25 & 10 & 2.96 & 5.4$\times 10^{-2}$ \\
SDSS J1204+1201 & $-4.34$ & $<+1.45$  & 10.6 & 0.9 & 1.41 & NA & 25 & 1 & 2.38 & 6.8$\times 10^{-2}$ \\
\multicolumn{11}{c}{[C/Fe]$>1$}\\
HE 2139-5432 & $-4.02$ & $+2.60$  & 28.0 & 0.6 & 9.51 & 3.36$\times 10^{-4}$ & 25 & 10 & 4.91 & 8.6$\times 10^{-4}$ \\
HE 2239-5019 & $-4.15$ & $+1.80$  & 15.0 & 10.0 & 1.43 & 1.06$\times 10^{-1}$ & 25 & 1 & 2.72 & 6.8$\times 10^{-2}$ \\
HE 1310-0536 & $-4.15$ & $+2.53$ &10.9 & 0.3 & 1.59 & 9.49$\times 10^{-3}$ & 15 & 1 & 2.86 & 1.7$\times 10^{-4}$ \\
CS 22949-037 & $-4.38$ & $+1.73$  & 27.0 & 0.3 & 12.03 & 2.01$\times 10^{-5}$ & 25 & 10 & 3.05 & 8.6$\times 10^{-2}$ \\
HE 0557-4840 & $-4.75$ & $+1.66$  & 10.9 & 0.6 & 1.41 & 1.28$\times 10^{-2}$ & 40 & 30 & 9.94 & 1.2$\times 10^{-1}$ \\
SDSS J1313-0019 & $-5.00$ &  $+2.96$ & 27.0 & 0.3 & 12.03 & 2.01$\times 10^{-5}$ & 25 & 1 & 5.36 & 8.5$\times 10^{-4}$ \\
\enddata
\tablenotetext{a}{Remnant mass for given progenitor mass and explosion energy taken from \citet{2008ApJ...679..639Z}}
\tablenotetext{b}{Ejected mass of $^{56}{\rm Ni}$ for given progenitor mass and explosion energy of the mixed model from \citet{2010ApJ...724..341H}}
\end{deluxetable*}

\subsection{Stars with large $\chi^2_{\nu}$ values}

In the present sample, the abundance fitting
for the eight objects results in $\chi^2_\nu > 8.5$. The
observed abundance ratios ([X/Fe]) and their best-fit models are shown in
Figure \ref{fig:outliers}. These stars can be broadly classified
into two categories based on their characteristic abundance ratios
as detailed below.

\subsubsection{Stars with a very low [Si/Fe] ratio}

\begin{figure}
  \plotone{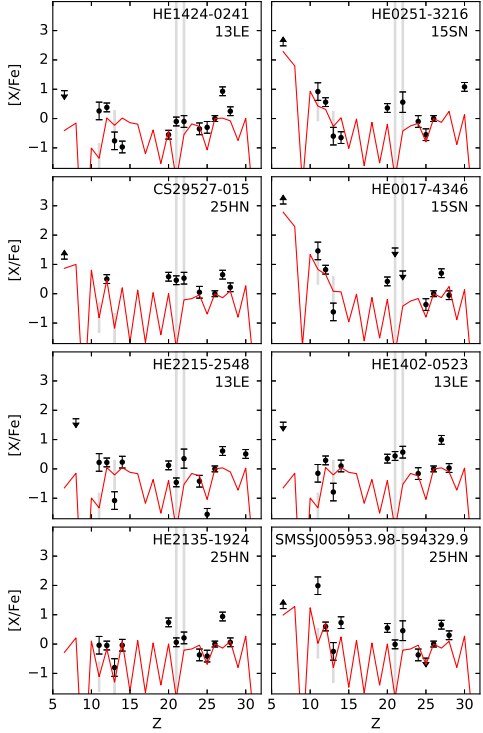}
  \caption{Model and observed abundance patterns for stars whose best-fit
    models have $\chi^2_{\nu}>8.5$. \label{fig:outliers}}
  \end{figure}

The top two panels in Figure \ref{fig:outliers} show the
observed abundances and the best-fit models for 
HE 1424-0241 and HE 0251-3216, both of which have extremely 
low [Si/Fe] ratios ($-1.0$ and $-0.7$,
respectively). Both objects show [Mg/Fe] ratios similar to other EMP
stars and thus [Si/Mg] ratios are very low. Despite the similarity
in [Si/Mg] between the two stars, their [(C+N)/Fe] and [Ca/Fe] ratios are
largely different; HE 1424-0241 shows a very low [Ca/Fe] ($\le-0.5$) ratio
 and there is 
 no sign of carbon enhancement \citep{2007ApJ...659L.161C,2013ApJ...778...56C}.
\citet{2009ApJ...690..526T} suggests that the abundance ratios of this star
are well reproduced with angle-delimited yields calculated
for a jet-induced supernova of a population III 40$M_\odot$ star. 
On the other hand, HD 0251-3216 is
a C-enhanced star with [C/Fe] ratio $\sim 2.5$ and show
a normal [Ca/Fe] ratio \citep{2013ApJ...778...56C}. Although there is no Sr measurements,
both [Y/Fe] and [Ba/Fe] ratios are relatively high ($>1$ dex) and thus
contribution from AGB nucleosynthesis cannot be ruled out.
For both stars, however, the low-[Si/Fe] ratios remain
challenging.

\subsubsection{Stars with very high [Co/Fe]}

Remaining six stars (CS 29527-015, HE 0017-4346,
HE 2215-2548, 
HE 1402-0523, HE 2135-1924, and SMSSJ005953.98-594329.9)
all show a very high [Co/Fe] ratio
($>0.5$) than
that typically observed in other EMP stars. The high [Co/Fe] ratio
is not attributed to the NLTE effects in the abundance analysis since 
the correction is positive and thus would further increases 
the discrepancy away from the theoretical yields \citep{2010MNRAS.401.1334B}.
These stars show a variety of abundances for the other elements. As an
example, HE 0017-4346
is a CEMP star with [(C+N)/Fe]$>1$ with an enhancement of
both [Na/Fe] and [Mg/Fe] ratios. The abundance pattern of
this star is marginally fitted with the 15SN model with
a small ejected mass of $^{56}$Ni. HE 2215-2548
shows enhancements of both [Co/Fe] and [Zn/Fe] ratios
($>0.5$ dex).  As discussed in
\citet{2009ApJ...690..526T}, high-entropy environment realized in
a simulation of
jet-induced supernova enhances Co and Zn but still
underestimate the observed abundances. 

\subsection{Hyper metal-poor stars with [Fe/H]$<-5$ \label{sec:HMP}}

The three most Fe-poor ([Fe/H]$<-5$) stars, HE 0107-4240 \citep{2002Natur.419..904C,2006ApJ...644L.121C}, HE 1327-2326 \citep{2005Natur.434..871F,2006ApJ...639..897A,2008ApJ...684..588F},
and SMSS 0313-6708 \citep{2014Natur.506..463K,2015ApJ...806L..16B,2017A&A...597A...6N}
are not included in the main sample because the numbers of reported
elemental abundances are small in the references used for
the other sample stars (see Section \ref{sec:obsdata}). For these three
stars, we use chemical abundances derived from 3D and/or NLTE analyses
when available mainly from \citet{2006ApJ...644L.121C,2008ApJ...684..588F,2017A&A...597A...6N}.
For these three stars, the abundance fitting method is applied by
treating C and N abundances separately with theoretical uncertainties
of 0.5 dex. Since [Fe/H] has not been
obtained for SMSS 0313-6708, the hydrogen mass is varied 
to reproduce the observed [Ca/H] abundances rather than [Fe/H]. 

\begin{figure}
  \plotone{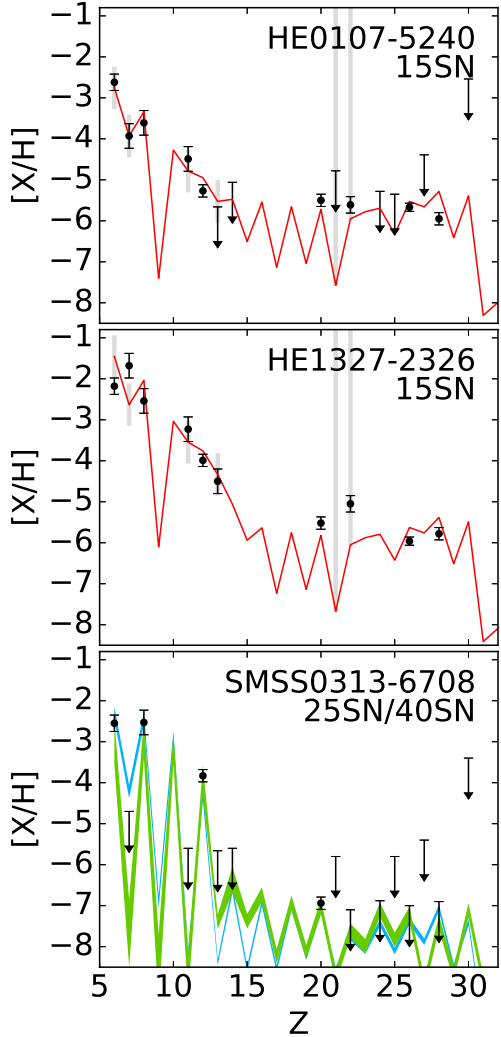}
  \caption{Best-fit models and observed abundances for
    the stars with [Fe/H]$<-5$. The top two
    panels show
    the best-fit 15SN models and the observed abundances
    for HE 0107-5240 and HE 1327-2326.
    The bottom panel shows observed abundances
    and upper limits for SMSS 0313-6708 and the ranges spanned by the
    models that fit the data with $\chi^{2}<10$, which are
    either the 25SN (blue band) or 40SN (green band) models.\label{fig:hmp_ump}}
  \end{figure}

The resulting best-fit models for HE 0107-5240 and HE 1327-2326 are
shown in the top and the middle panels of Figure \ref{fig:hmp_ump}.
The abundances of both stars are best-fitted with the
model for a Pop III star with $M=15M_{\odot}$ which explodes with a normal
explosion energy ($E_{51}=1$). The best-fit mixing-fallback model
parameters for these two stars suggest that they leave behind a compact
remnant with $M=2.9M_{\odot}$ and eject a very small amount of
$^{56}$Ni ($<10^{-4} M_\odot$). 

For SMSS 0313-6708, abundance measurements for only C, O, Mg, and Ca
are available, while upper limits for other elements have been obtained
either from 1D/3D LTE or 3D/NLTE analyses \citep{2015ApJ...806L..16B,2017A&A...597A...6N}.
The ranges spanned by the models with $\chi^2$ smaller than 10 are shown
in the bottom panel of Figure \ref{fig:hmp_ump}. As the result of the
abundance fitting, only the 25SN ($M=25M_{\odot}$ and $E_{51}=1$; blue band) or
40SN ($M=40M_{\odot}$ and $E_{51}=1$; green band) models fit the data with $\chi^2<10$.
Among them, the low upper limit for the [N/H]
abundance is consistent with the 40SN model. 
The progenitor mass of $40M_\odot$ is similar to that suggested by \citet{2015ApJ...806L..16B}
based on the {\it STARFIT} code \citep{2010ApJ...724..341H}; 
  the model for a $40 M_{\odot}$ Pop III star which explodes with $E_{51}=1.8$
  with modest mixing explain the observed abundance pattern.
  The origin of Ca in the models of \citet{2015ApJ...806L..16B} and
  this work, however, is different; namely, Ca in the model
  of \citet{2015ApJ...806L..16B} is produced in the outer layer by 
  the hot-CNO cycle during the 
  pre-supernova evolution while it is produced by
  static/explosive O and Si burning in the model presented in this work
  \citep{2014ApJ...792L..32I}.
  While the new abundance constraints from 3D-NLTE abundance analysis
  by \citet{2017A&A...597A...6N} are consistent with the latter
  scenario as shown in the bottom panel of
  Figure \ref{fig:hmp_ump}, additional abundance measurements,
  especially for Fe-peak elements, are necessary to distinguish
  between the two Ca production scenarios. 
  Since the hot-CNO cycle
  can occur only in a zero-metallicity star,
  the origin of Ca in SMSS 0324-6708 could be an important
  diagnostics to examine whether or not the progenitor
  of this star has to be a Pop III star.

\section{Discussion} \label{sec:discussion}

\subsection{Implications on the initial mass function of the Pop III stars}
One of the major goals of this study is to
  derive the IMF of the first stars from the observed
  elemental abundances of EMP stars. We note, however, that the
  mass function we obtain in this abundance fitting
  analysis is the IMF of the first metal-enriching stars, which
  are not nessesarily the same as the IMF
  of the first stars. 
  In Section \ref{sec:PopIIImass}, we show that, among the adopted
Pop III supernova models (Low-energy explosion for 13$M_\odot$,
SNe for 13, 15, 25, 40, 100$M_{\odot}$ and
HNe for 25, 40, and 100$M_\odot$ Pop III stars), the majority
of the EMP stars with available abundance measurements are best explained
by the models of Pop III stars with masses $15$ and $25 M_{\odot}$
(Figure \ref{fig:mhist_SNHN}). The mass function of the first metal-enriching
stars is well represented by a log-normal function, \begin{equation} \propto \exp(-(\ln x-\mu)^2/2\sigma^2) \end{equation} with $(\mu,\sigma)=(3.28\pm0.02,0.31\pm0.01)$ for the non-weighted histogram and $(3.30\pm 0.03,0.36\pm 0.02)$ for the weighted histogram.

More specifically,
  at $M=13$ or $15M_{\odot}$, the Pop III models fitting observed abundances
  are $\sim 50$\% less frequent than at $M=25M_\odot$. This is different
  from the Salpeter
  IMF, which has a power-low form of $M^{-2.35}$,
  as shown in the dotted lines in Figure \ref{fig:mhist_SNHN}. 
  The NLTE correction
to the Al abundances increases the contribution from the $M=13$ and $15M_\odot$
models but the $M=25 M_\odot$ model remains dominant in
the histogram as shown in Figure \ref{fig:mhist_systematics}.

  Similarly, at $M\ge 40M_{\odot}$, the best-fit Pop III models are about one third of those of the $M=25M_{\odot}$.
  The smaller contribution from the larger Pop III mass ($M\geq 40 M_\odot$)
implies that either (1) the formation mechanism of the first
stars inhibit the formation of $\gtrsim 40 M_{\odot}$ stars, (2)
the first stars with $\gtrsim 40 M_{\odot}$ directly
collapse into a black hole remnant without ejecting any nucleosynthetic
products or (3) the supernova explosions of higher-mass first stars inhibit
  the formation of the next-generation stars \citep[e.g.,][]{2014ApJ...791..116C}.

  \citet{2013MNRAS.433.1094S} took into account the formation of multiple
    stellar systems in their cosmological simulation and found that
    the maximum Pop III mass is limited to $M\sim 40M_{\odot}$. This mass range
    is consistent with the scenario (1), where the formation of
    $M\ge 40M_{\odot}$ Pop III stars is inhibited. 
    On the other hand, \citet{2014ApJ...781...60H} predict
    abundant formation of more massive stars ranges from $M=10M_{\odot}$ up to
    a few thousands of $M_{\odot}$ by taking into account
statistical variation of the properties of primordial star-forming clouds 
in a cosmological context \citep[see also][]{2017MNRAS.470..898H}. Using three-dimensional radiation
hydrodynamics simulations, \citet{2014ApJ...792...32S} also predicts a similar
but slightly lower mass range for the Pop III stars
($1\lesssim M\lesssim 300 M_\odot$). These two results are more in line with
the scenario (2), where the Pop III stars with $M\ge 40M_{\odot}$
can be formed but do not eject any heavy elements
via their supernova explosions.  

It is shown by theoretical studies
that the progenitors more massive than $\sim 40-50M_\odot$ more likely
collapse to form black holes without explosion
\citep[e.g.,][]{1999ApJ...522..413F,2002ApJ...567..532H}.
\citet{2012ApJ...748...42C} also predict
that Pop III stars with mass as low as $M\sim 80M_\odot$ can  explodes as 
pair-instability supernovae rather than core-collapse
supernovae depending on rotation of the progenitor star.
The inffered scarcity of chemical signature from $M\ge 40M_\odot$ is
therefore consistent with these theoretical expectation.

\begin{figure*}
  \plottwo{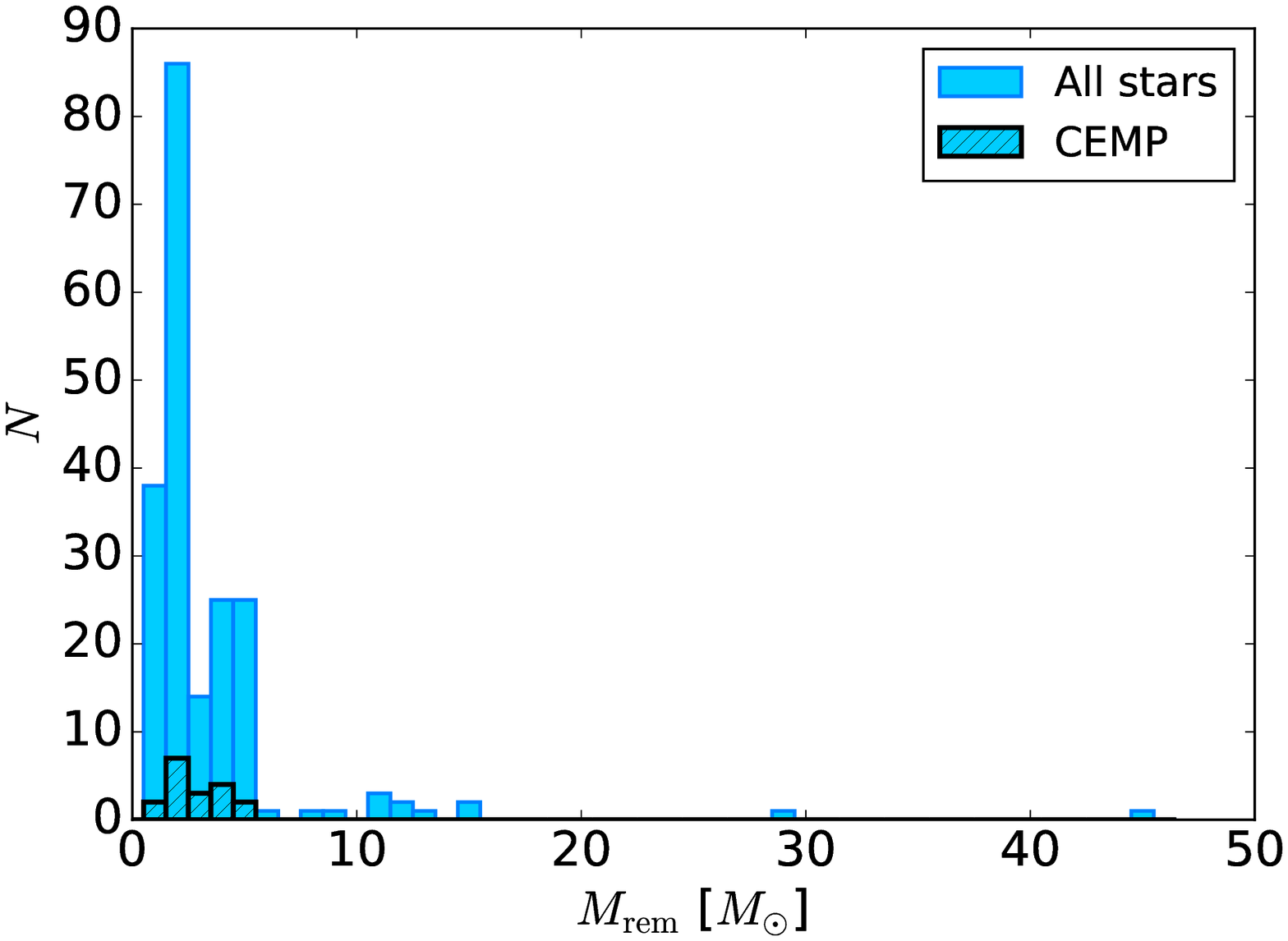}{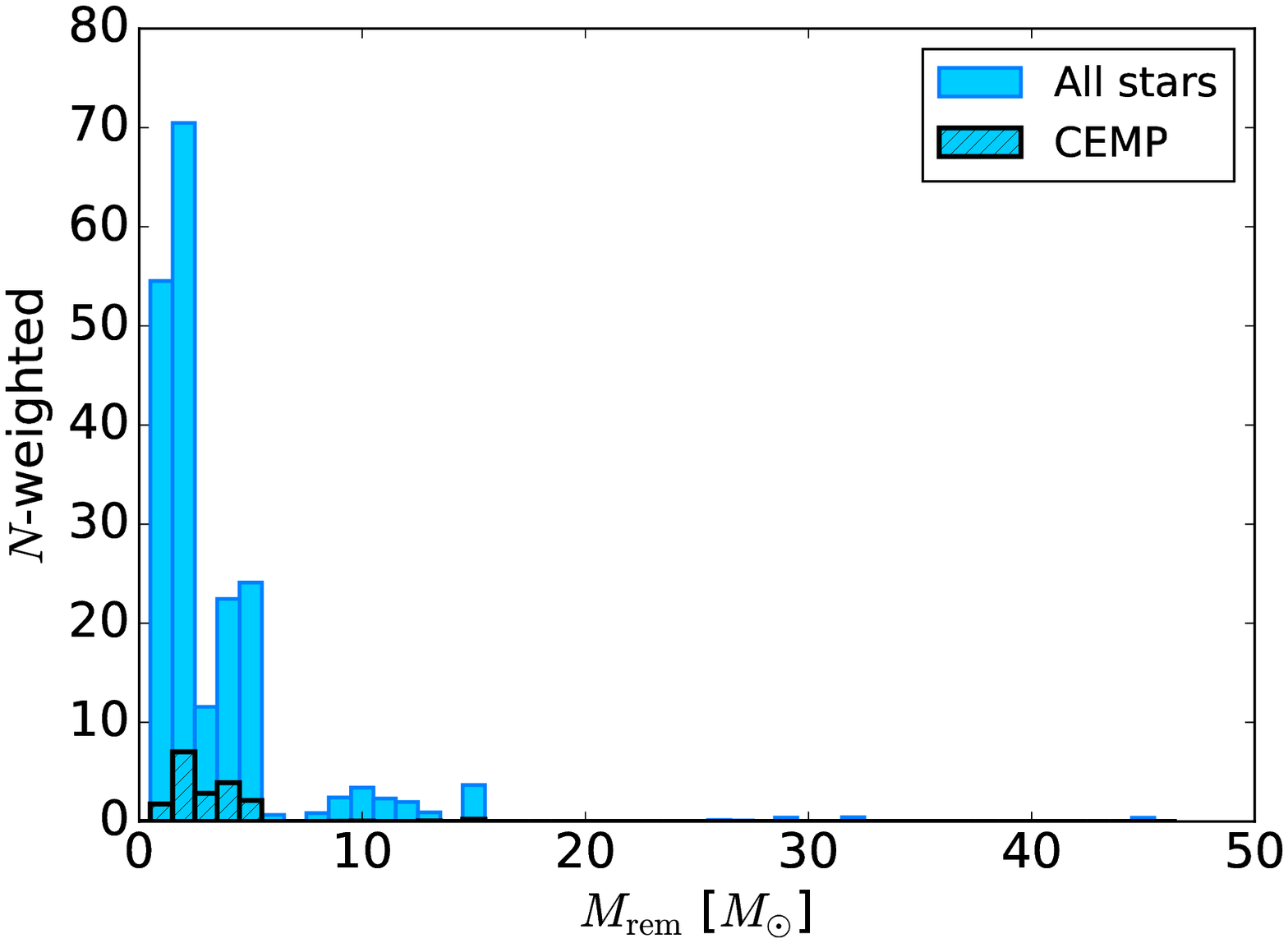}
  \caption{{\it Left:} a histogram of the remnant mass $M_{\rm rem}$ obtained
      by Equation \ref{eq:mrem} from the best-fit parameters to each stars.  
      {\it Right:} the corresponding histogram obtained by counting contributions
      from all the mass-energy models weighted by the p-values. The hatched
    histograms correspond to the histograms for the CEMP stars \label{fig:Mrem}}
\end{figure*}

  Observational constraints on the masses of the stellar mass blackholes
  that are possible first-star remnants are helpful to distinguish between
  the scenarios (1) and (2)  \citep[e.g.,][]{2010ApJ...725.1918O,2016PhRvL.116f1102A,2016MNRAS.456.1093K,2016MNRAS.460L..74H}, which are complementary to the chemical signature of EMP stars. 
  For example, in the case of the scenario (1),
  the blackhole mass distribution of the first stars
  should be the same as the compact remnant mass distribution
  obtained by Equation \ref{eq:mrem} in this work.
   Figure \ref{fig:Mrem} shows the distribution of the masses of the
  compact remnant for the best-fit models (left) and that obtained by
weighting with the $p-$values (right).
The distribution 
is predominantly peaked at $\sim 1.5-3M_{\odot}$ with a tail
extending to  $\sim 46M_\odot$.
In the scenario (2), on the other hand, masses of the compact remnant
  could be much larger than those shown in Figure \ref{fig:Mrem}.
  Since mass loss is expected to be negligible
for Pop III stars because of the low opacity in their atmosphere, which
prevent strong stellar winds, the final mass of the Pop III is likely
preserve its original mass, which may 
finally collapse without ejecting synthesized elements.

In the scenario (3), the elements synthesized by Pop III stars in the
range 40-100$M_\odot$ are ejected to inter-galactic medium
but low-mass stars do not form
out of gas containing the ejecta, which may not satisfy
various physical conditions required for low-mass star formation
\citep[e.g.,][]{2015MNRAS.452.2822S}. The ejected elements would have
contributed to the inter-galactic medium and thus elemental abundance
signature of the $>40M_\odot$ Pop III stars would remain
in gas-phase metals in high-redshift objects. 
Measurements of gas-phase metallicity for high-redshift
objects such as Damped Lyman-Alpha systems would be useful to test
this scenario \citep{2011ApJ...730L..14K,2017MNRAS.467..802C}.

  \subsection{Explosion energies}
  Another important finding of our analysis is for the explosion energies
  of the Pop III core-collapse supernovae; almost half of the sample 
  stars are best-fitted with the model for a Pop III star with 
  $M=25M_{\odot}$ which explodes with high explosion energy ($E_{51}=10$).
  Such a large fraction of hypernovae might be responsible for the
  chemical evolution of the Milky Way,
  not only in the solar neighborhood \citep[e.g.][]{2006ApJ...653.1145K,2010A&A...522A..32R,2011ApJ...729...16K},
  but also in the Galactic bulge \citep{2015Natur.527..484H}.  
  At low metallicity, [Zn/Fe] ratios show an increasing trend toward lower metallicities
  \citep{2000fist.conf...51P,2004A&A...416.1117C}, which can be reproduced with hypernovae \citep{2005ApJ...619..427U}.
  
  In nearby universe, the most likely progenitors of hypernovae
  are thought to be rapidly rotating He cores that have stripped their
  H envelope that end up with energetic Type Ibc supernovae and they
  are rare compared to typical Type II SNe \citep[e.g.][]{2004ApJ...607L..17P}.
  The rotational properties of the present-day and Pop III supernova
  progenitors should be very different  
  because of the lack of mass loss due to the low-opacity, which prevent the star to loose the angular momentum. Note that, however, it is still debated that whether the Pop III stars can
  maintain the high rotational velocity under the 
presence of magnetic fields \citep[e.g.][]{2012A&A...542A.113Y,2016A&A...585A.151L}.

At this moment, the only observational signatures of hypernovae among
Pop III stars come from EMP stars, and there is no complementary observational
evidence to support the high fraction of hypernovae in the early universe.
The direct detection of light curves of Pop III SNe \citep[e.g.][]{2014ApJ...797...97S,2016ApJ...821..124T}
is necessary to obtain more robust insights into the nature and the explosion
mechanisms of the Pop III stars, which would eventually better constrain
the Pop III IMF.
That would require the detection of bright supernovae at high-redshifts by
next-generation instruments
such as the {\it Wide-Field Infrared Survey Telescope} (WFIRST), the {\it Large Synoptic Survey Telescope} (LSST) or the {\it James Webb Space Telescope} (JWST) \citep{2017arXiv171105742H}.

\subsection{Ejected mass of $^{56}$Ni}

\begin{figure*}
  \plottwo{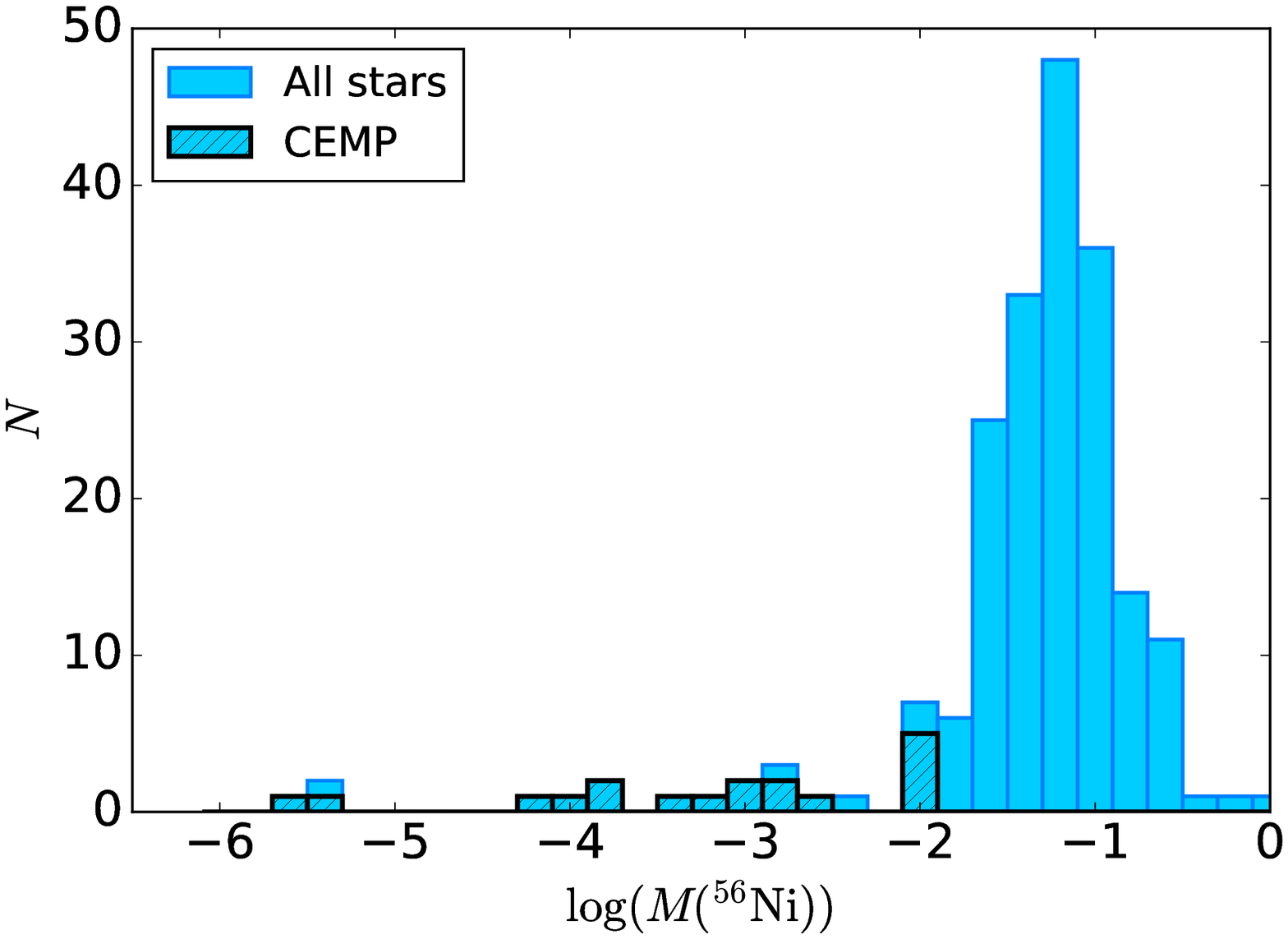}{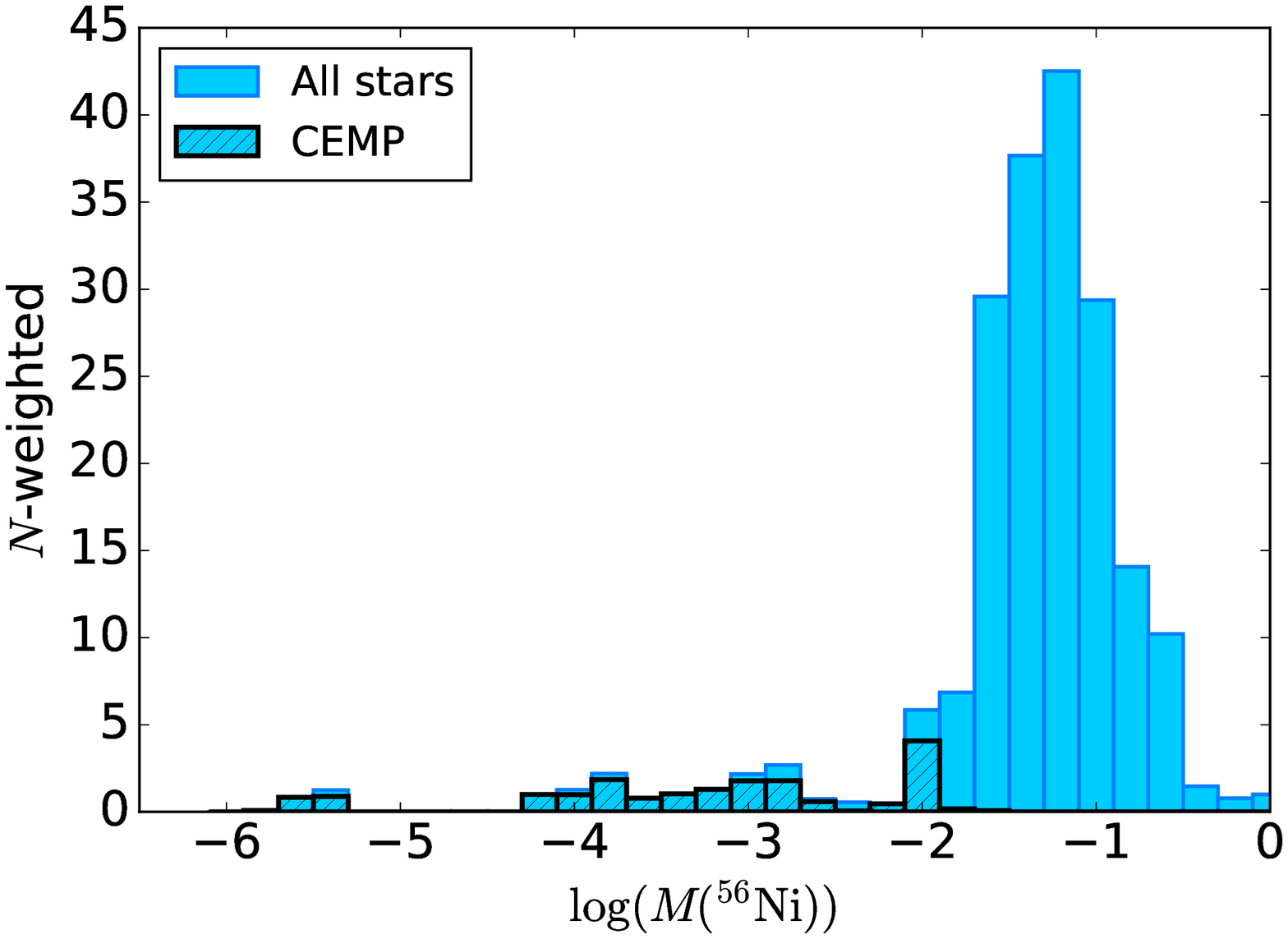}
  \caption{Similar histograms as in Figure \ref{fig:Mrem} but for the ejected
    mass of $^{56}$Ni. \label{fig:Mni}}
\end{figure*}

The yield of radio-active $^{56}$Ni is one of the main source of
luminosity in supernovae. The ejected $^{56}$Ni mass is measured by
multi-color light-curve analyses of local supernovae. Also, this
isotope finally decays to $^{56}$Fe, the primary
stable isotope of Fe.
Figure \ref{fig:Mni} shows a histogram of ejected mass of the $^{56}$Ni
from the best-fit models.
It can be seen from both of the direct count of the best-fit models (left)
and the $\chi^{2}_{\nu}$-weighted count that the ejected
$^{56}$Ni mass is predominantly
0.01-0.1$M_\odot$. These masses is similar to those estimated by
light-curve analysis of local supernova observations \citep[e.g.,][]{2017ApJ...841..127M}. 
On the other hand, the low-$^{56}$Ni mass tail
is very different from the distribution of the
rest of the sample, as the tail is caused from the existence of
CEMP stars. A fraction of objects that are best fitted with
models that eject only small amount of $M(^{56}$Ni$)<0.01 M_\odot$,
which presumably corresponds to faint supernovae, is $\sim 10$
\% of the first supernovae.
Whether or not the typical ejected $^{56}$Ni masses and
the fraction of faint supernovae
implied by the present analysis is consistent with the amount of metals in
present-day galaxies should be tested through chemical evolution models
with our yield of faint supernovae ($<10^{-3} M_{\odot}$).

\subsection{Element ratios as the mass indicator \label{sec:abratios}}

Through our analysis, we can also propose
the mass indicator of element ratios. 
In the previous sections, we obtained the best-fit model parameters
to each EMP star. Based on these best-fit models, we investigate which 
abundance ratios best correlate with progenitor masses.

\begin{figure*}
  \plottwo{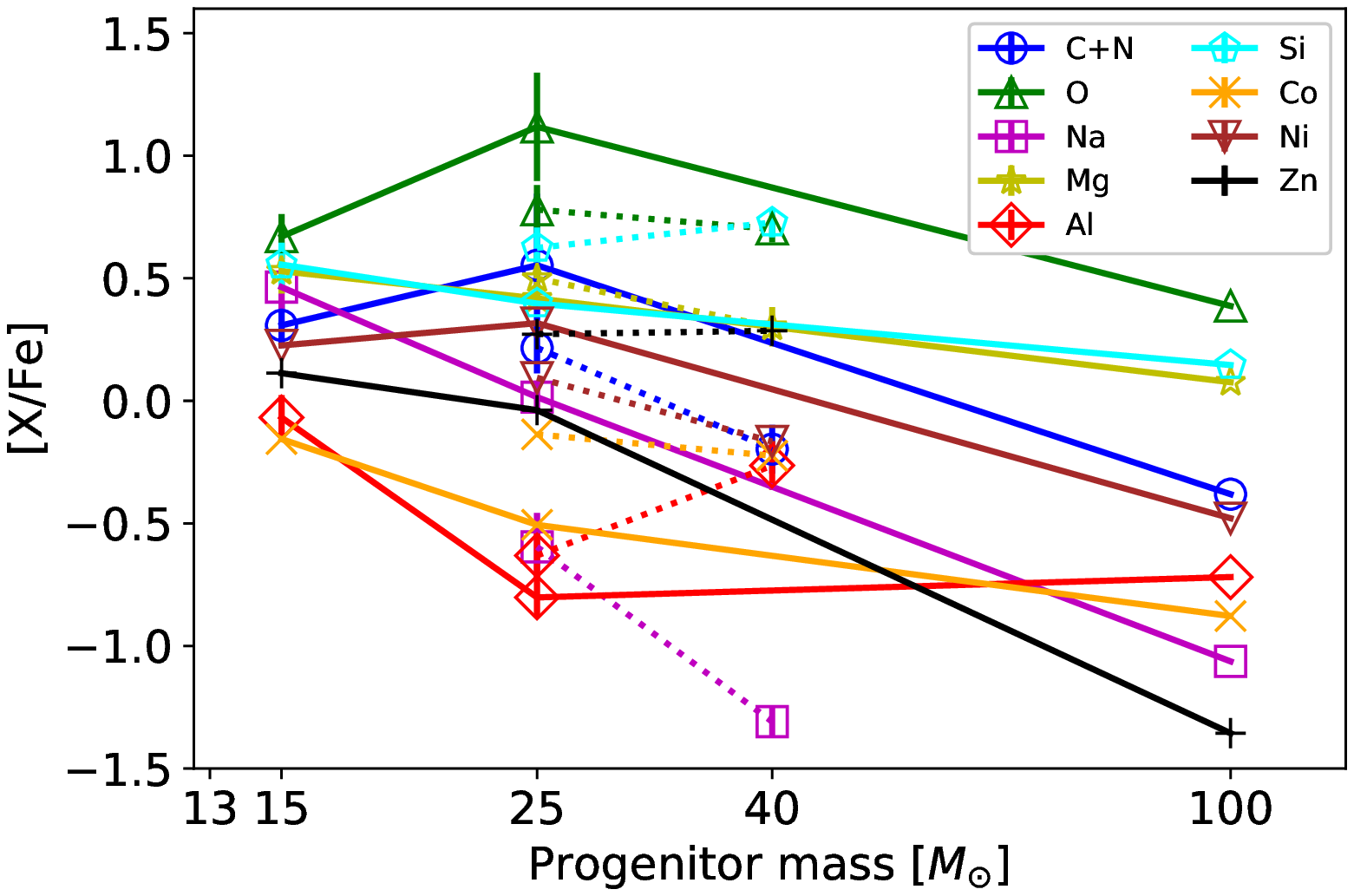}{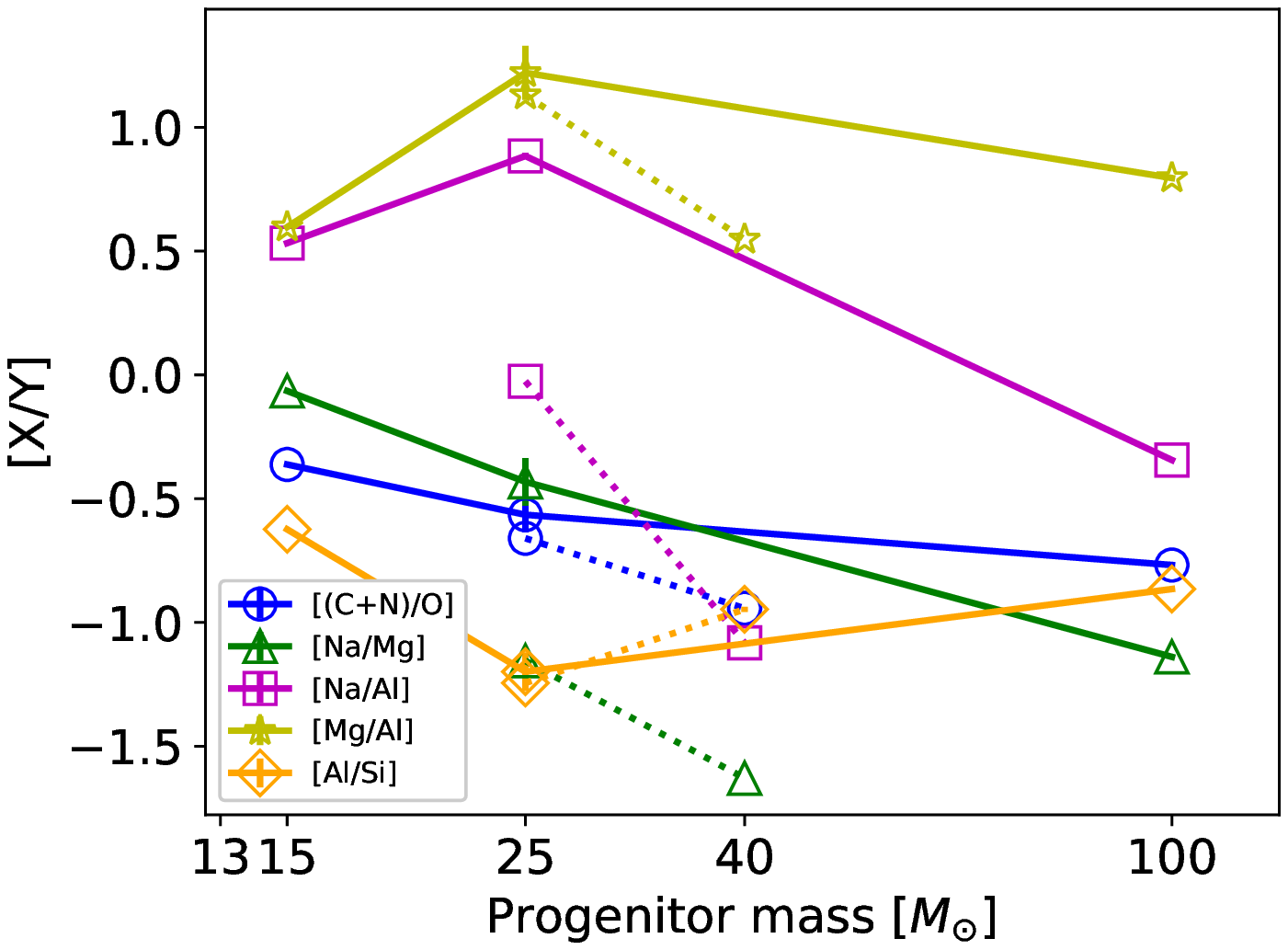}
  \caption{Abundance ratios as a function of masses of the Pop III stars for the best-fit models. The left panel is for the ratios relative to Fe and the right panel shows the ratios among C+N, O, Na, Mg, Al, and Si, for which abundance ratios are less affected by the choice of the mixing-fallback parameters. \label{fig:mass_ratios}}
  \end{figure*}

Figure \ref{fig:mass_ratios} shows
the median abundance ratios of the best-fit models (
15SN, 25SN, 25HN, 40HN and 100SN models) plotted against the progenitor masses
(i.e., either 15, 25, 40 or 100 $M_{\odot}$).
The left and right panels
show abundance ratios relative to Fe and the ratios among the light elements,
respectively. The solid and dotted lines
indicate the trends for the supernova (15SN/25SN/100SN) models and the
hypernova (25HN/40HN) models, respectively. The error bars in these plots
represent the median absolute deviation of the best-fit
models of a given progenitor mass/explosion energy.

As can be seen in the left panel, the [Na/Fe], [Mg/Fe], [Si/Fe],
[Co/Fe] and [Zn/Fe] ratios monotonically decrease 
with the progenitor mass for the SN models (solid lines). 
These ratios, however, are not necessarily
correlate with the mass for the HN models (dotted lines).
This result highlights the importance of constraining
the explosion energies from multiple abundance measurements
including e.g., a [Zn/Fe] ratio.

Among the light elements, shown in the right panel,
[(C+N)/O] and [Na/Mg] ratios
have negative correlation with progenitor masses for both of the
SN and HN models.
The decreasing trend of the [(C+N)/O] ratio with increasing
the Pop III main-sequence masses
stems from the fact that C is mainly synthesized in
the C+O layer between the He layer and the convective O core, of which
temperature is moderately high to ignite the He burning but not the C
burning. Since production of O more strongly depends on
  temperature, and hence main-sequence masses,
the C mass increases more slowly than the O mass
with the main-sequence masses.
The trend of [Na/Mg] resulted from the fact that 
synthesis of Na is less efficient
in the C shell burning with higher temperature, which is typically realized
for more massive Pop III stars.

\begin{figure}
\plotone{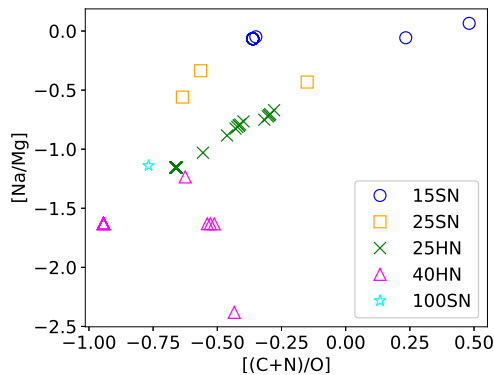}
\caption{[(C+N)/O] vs [Na/Mg] ratios for the models that fit the EMP stars. \label{fig:cno_namg}}
\end{figure}

To summarize, among the best-fit models considered
in this work, the elements that are sensitive to the progenitor
masses are the ratios between C$+$N, O, Na, and Mg. 
Figure \ref{fig:cno_namg} summarize the locations in the [(C+N)/O] vs
[Na/Mg] ratios for the best-fit models.
The figure shows that a star with [Na/Mg]$>-1.0$ and [(C+N)/O]$>-0.6$ is
more likely to be fitted with either the 15SN, 25SN, or 25HN models
while a star with [Na/Mg]$<-1.0$ and with [(C+N)/O]$<-0.4$ is
more likely to be fitted with either the 40HN or 100SN models.

The correlation of the C/O ratio with the progenitor
masses is expected from stellar evolution theories.
The ratio in supernova ejecta, however,
could significantly depend on the mixing-fallback process. 
At the same time, Na is burnt to Mg and Al in explosive nucleosynthesis
and thus the Na/Mg ratio depends on not only Pop III mass but
the explosion energies. 
We, therefore, emphasize that multiple elemental
abundance measurements, other than the [(C+N)/O] and [Na/Mg] ratios,
are also essential to resolve the degeneracy among the mixing-fallback
process, explosion energies and the masses of the Pop III supernovae.

\subsection{Limitation of the present approach}
\label{sec:limitation}

\subsubsection{Uncertainties in stellar evolution}

One of the major limitations in this study is that
the Pop III model employed 
does not include the effect of stellar rotation.
The rotation has been suggested to play an important role in
determining the structure and the nucleosynthetic yields of
massive stars 
\citep{2001A&A...373..555M,2007A&A...461..571H,2010ApJ...723..353J}. 
In a rotating Pop III stellar evolution model,
production of $^{14}$N is enhanced
as the result of $^{12}$C mixed from the helium-burning shell
into the hydrogen-burning shell, in which the
CNO cycle takes place \citep{2002A&A...381L..25M}. Consequently,
intermediate-mass isotopes such as $^{23}$Na or $^{27}$Al are also
produced through a series of alpha capture reactions on
$^{14}$N \citep{2014ApJ...794...40T}. Therefore, the abundance ratios
involving Na or Al may not clearly correlate with the
Pop III progenitor mass if the star had rotation.  

The prediction on these odd-Z elements are also
affected by the treatment of overshooting or the
uncertainty in the $^{12}$C($\alpha$,$\gamma$)$^{16}$O reaction rate
\citep{2002ApJ...577..281C}. 
Therefore, the prediction on Na and Al, in particular, as a Pop III
mass indicator should be viewed with caution.

\subsubsection{Explosive nucleosynthesis in multi-dimensional simulation}

In this work, we utilize the mixing-fallback model applied to
the one-dimensional nucleosynthesis calculation to approximate
the Pop III yields of aspherical supernovae. An important issue to be
verified is
that the departure of the calculated yields from those predicted
by multi-dimensional simulations of aspherical supernovae. 

\citet{2009ApJ...690..526T} performs a two-dimensional hydrodynamical
and nucleosynthesis calculation for an aspherical jet-induced explosion of a
40$M_{\odot}$ Pop III star. The results suggest 
that the angle-averaged yields for many elements
are in agreement with those from the particular parametrization in the
mixing-fallback model applied to the one-dimensional nucleosynthesis
calculation. However, elemental ratios such as
[Sc, Ti, V, Cr, Co, Zn/Fe] are enhanced in the simulation
as the result of the high-entropy environment realized only
in the two-dimensional calculation of the jet-induced supernovae.
It is also demonstrated that the predicted yields largely depend 
on the angle from the jet axis and thus if the ejecta is
not well mixed, the elemental abundances imprinted on the next
generation of stars could be significantly different from
those predicted by the mixing-fallback model 
\citep{2009ApJ...690..526T}.

In order to better constrain the Pop III progenitor of EMP stars
based on the measurements of Sc, Ti, V, Cr, Co, and Zn abundances,
multi-dimensional simulations with various progenitor masses
and explosion energies are needed.

\section{Conclusion}
\label{sec:conclusion}

We calculate supernova yields of Pop III stars in the mass range 13-100$M_{\odot}$
for low ($E_{51}<1$), normal ($E_{51}=1$) and high ($E_{51}>1$) explosion energies 
that best reproduce elemental abundance measurements of
$\sim 200$ EMP stars taken from recent literature in the framework
of the mixing-fallback model.
The results can be summarized as follows.

\begin{itemize}
\item{The observed abundances of the majority of the
  present sample of EMP stars are best-reproduced
  with the Pop III yields from progenitors being
  less massive than 40$M_{\odot}$. Almost half
  of the sample stars are best-fitted with the model for a Pop III star with 
  $M=25M_{\odot}$ which explodes with high explosion energy ($E_{51}=10$). }
\item{The predominance of the $M<40M_{\odot}$ best-fit Pop III models  
  is affected by the fiducial observational errors we have assigned
  to the data (0.1-0.3 dex). Obtaining a tighter constraint on the
  Pop III masses 
  requires the errors to be smaller than the fiducial values.} 
  \item{ We have also examined the effects of the
  two major systematic uncertainties; (1) NLTE effects on
  Al abundances and (2) the effect of non-measurements for
  specific elements.
  For (1), the
  uniform NLTE correction for the Al abundance by 0.6 dex results in the
  change in the best-fit models from $M=25M_{\odot}$ to $M\le 15M_{\odot}$
  for some of our sample stars,
  which highlights the necessity of the NLTE abundance measurements
  to discriminate between these progenitor masses.
  For (2), the lack of
  either the Si or Zn measurements leads to the change in the best-fit
  Pop III models and thus measurement of these elements is
  particularly important. 
  In both of (1) and (2), the main results, namely,
    the Pop III mass distribution
  is peaked at $M=25M_\odot$ and 
  the dominant contribution comes from $M<40M_{\odot}$, remain unchanged. }
  \item{The mixing-fallback parameters for most of
    the EMP stars
    are characterized by (1)$M_{\rm mix}$ smaller than the mass below which
  the explosive nucleosynthesis take place and (2)$f_{\rm ej} \sim 0.01-0.5$. 
  The results indicate that the progenitor Pop III super-/hypernovae have predominantly
  left behind the
  compact remnants with masses less than $<5M_{\odot}$ and ejected $\sim 0.01-0.1M_{\odot}$ of
  $^{56}{\rm Ni}$.}
\item{The CEMP stars ([(C+N)/Fe]$>1.0$) in our sample are best-fitted
  with the Pop III models with progenitor masses similar to those fit the other C-normal EMP stars.
   On the other hand, the best-fit mixing-fallback parameters for the CEMP stars are largely different from those of the majority of other EMP stars in that 
   the CEMP are explained by the large mixing region and the small ejected fraction. The CEMP stars with Mg enhancement ([Mg/Fe]$>1$) are explained by the model for a Pop III star with $M=25M_{\odot}$ exploding with the high explosion energy (hypernova). }
\item{The resulting mass distribution of the progenitor Pop III of the EMP
  stars decreases at $M<25M_{\odot}$, which is not consistent with
  the Salpeter IMF. The drop at $M\ge40_{\odot}$ suggests either that 
  (1) the formation of the first stars with  $M\ge 40M_{\odot}$ are suppressed, that 
  (2) the $M\ge 40M_{\odot}$ first stars tend to directly collapse into black holes without ejecting any heavy elements to be
  incorporated into the next generation of low-mass stars, or that (3) the supernovae of higher-mass Pop III stars inhibit the formation of the next-generation of low-mass stars. These scenarios predict different
  distributions of mass of the compact remnants and ejected mass of $^{56}$Ni and thus
  should be tested with other observational probes such as masses of the stellar-mass
  black holes and light-curves of Pop III supernovae in the future observations. }
\item{Based on the best-fit models, we propose the diagnostic abundance ratios sensitive
  to the Pop III progenitor masses, where the [(C+N)/O] ratios best correlate with the progenitor masses. The Na, Mg, and Al abundances could also be
  sensitive to the progenitor masses if the progenitor
  stellar rotation does not significantly affect the abundances of these elements.
  }
\end{itemize}

These results demonstrate that the elemental abundances in
EMP stars have useful implications on the 
physical properties of the Pop III stars and  
their supernova explosions. On-going and future 
photometric and spectroscopic surveys such as the
SDSS/APOGEE \citep{2017AJ....154...94M}, GALAH \citep{2015MNRAS.449.2604D},
Pristine survey \citep{2017MNRAS.471.2587S}, LAMOST \citep{2012RAA....12..723Z}, and their
follow-up observations \citep[e.g.][]{2015PASJ...67...84L} to
accurately measure most of the important key elements for large
samples of EMP stars are crucial to obtain more robust
insights into the nature of the Pop III stars. 

The present analysis method, however, is based on the various
assumptions on the progenitor Pop III stellar evolution (rotation) and supernova explosions,
which should be verified with more realistic
multi-dimensional nucleosynthesis calculations. At the same time,
the present results highlight the importance of 
complementary high-redshift supernova observations with the
next-generation photometric and spectroscopic facilities \citep[e.g. {\it WFIRST}, {\it LSST},
and {\it JWST}; see][]{2017arXiv171105742H} to connect the nucleosynthetic signatures on EMP
stars with the initial mass function of the Pop III stars. 

\acknowledgments
M. N. I. thank A. Tolstov, S-C. Leung, A. Zhiglo and T. Hartwig
for fruitful conversations
on theoretical aspects of supernovae, nucleosynthesis and/or cosmological
simulations. The authors are grateful to N. Christlieb, A. Frebel and M. Limongi for helpful comments and suggestions. C.K. thanks Marii Shirouzu and Natsuko Izutani for
the earlier attempts on this topic.
This work has been supported by the World Premier International
Research Center Initiative (WPI Initiative), MEXT, Japan, and JSPS
KAKENHI Grant Numbers JP26400222, JP16H02168, JP17K05382, JP17K14249,
and the Endowed Research Unit "Dark side of the Universe" by Hamamatsu
Photonics K.K. at Kavli IPMU.

\vspace{5mm}

\appendix

\section{Table of best-fit models}

Table \ref{tab:bestfitmodels} presents [X/H] abundances of
the best-fit models and the observational data from literature
used in the abundance fitting.

\begin{deluxetable*}{lccccccccccccc}
\tablecaption{Best-fit models and observed abundances from literature\label{tab:bestfitmodels}}
\tablehead{
  \colhead{Starname} & \colhead{$M$} & \colhead{$E$} & \colhead{$M_{\rm mix}$} & \colhead{$\log f_{\rm ej}$} & \colhead{[C/H]$_{\rm mod}$} & 
  \colhead{flag \tablenotemark{a}} & \colhead{[N/H]$_{\rm mod}$} & \colhead{...} & \colhead{[C/H]$_{\rm obs}$} & \colhead{flag\tablenotemark{a}} & \colhead{[N/H]$_{\rm obs}$}&  \colhead{...} & \colhead{Ref.\tablenotemark{b}}
}
\startdata
HE0020-1741&  13&  0.5&  1.7&  $-1.1$&  $-3.49$&  1& $-4.57$ &   ... & $-2.24$&  1&  & ...&  9
\enddata
\tablenotetext{a}{Flags for the abundances: 1 for used value, -1 for upper limit, -2 for lower limit.}
\tablenotetext{b}{List of reference: (1) \citet{2013ApJ...762...26Y}, (2) \citet{2013ApJ...778...56C}, (3) \citet{2014AJ....147..136R}, (4) \citet{2015ApJ...807..171J}, (5) \citet{2014ApJ...787..162H}, (6) \citet{2015ApJ...809..136P}, (7) \citet{2015ApJ...810L..27F}, (8) \citet{2016AA...585L...5M}, (9) \citet{2016ApJ...833...21P}}
\tablecomments{Table \ref{tab:bestfitmodels} is published in its entirety in the electronic edition of the Astrophysical Journal. A portion is shown here for guidance regarding its form and content.}  
\end{deluxetable*}

\bibliography{mi21_astroph}



\end{document}